**Sepsis is a syndrome with hyperactivity of TH17-like innate immunity and hypoactivity of adaptive immunity**

By Wan-Chung Hu*
*Postdoctorate
Genomics Research Center
Academia Sinica
No 128 Academia Road section2
Nangang 115, Taipei, Taiwan

**Current Institutes:**

Department of Neurology
Shin Kong Memorial Hospital
Taipei, Taiwan

**Abstract**

Currently, there are two major theories for the pathogenesis of sepsis: hyperimmune and hypoimmune. Hyperimmune theory suggests that cytokine storm causes the symptoms of sepsis. On the contrary, hypoimmune theory suggests that immunosuppression causes the manifestations of sepsis. By using microarray study, this study implies that hyperactivity of TH17-like innate immunity and failure of adaptive immunity are noted in sepsis patients. I find out that innate immunity related genes are significantly up-regulated including CD14, TLR1,2,4,5,8, HSP70, CEBP proteins, AP1(JUNB, FOSL2), TGF-β, IL-6, TGF-α, CSF2 receptor, TNFRSF1A, S100A binding proteins, CCR2, formyl peptide receptor2, amyloid proteins, pentraxin, defensins, CLEC5A, whole complement machinery, CPD, NCF, MMP, neutrophil elastase, caspases, IgG and IgA Fc receptors(CD64, CD32), ALOX5, PTGS, LTB4R, LTA4H, and ICAM1. Majority of adaptive immunity genes are down-regulated including MHC related genes, TCR genes, granzymes/perforin, CD40, CD8, CD3, TCR signaling, BCR signaling, T & B cell specific transcription factors, NK killer receptors, and TH17 helper specific transcription factors(STAT3, RORA, REL). In addition, Treg related genes are up-regulated including TGFβ, IL-15, STAT5B, SMAD2/4, CD36, and thrombospondin. Thus, both hyperimmune and hypoimmune play important roles in the pathophysiology of sepsis.

## Introduction

Despite of the discovery of antibiotics, mortality rate of sepsis is still very high. Most important of all, the exact pathophysiology of sepsis is still unclear. Currently, there are two dominant theory to explain the etiology of sepsis: hyperimmune theory and hypoimmune theory. However, these two theories are contrary with each other. Hyperimmune theory was proposed by Dr. Lewis Thomas. In his classical paper in NEJM 1972, he proposed that hyperactivation of proinflammatory cytokines, the cytokine storm, is the actual cause of sepsis symptoms. These uncontrolled cytokines destruct and cause multiple organ failure. His theory is the mainstream theory of sepsis etiology. Based on this theory, therapeutic strategy such as antibody neutralizing TNFα was tested in septic patients in clinical trials. However, these antibodies did not improve the survival rate of septic patients. Further, anti-TNFα increased the mortality rate of septic patients in several clinical trials. That makes people to doubt the hyperimmune theory. Thus, another theory-hypoimmune theory emerges. Based on the observation that immunosuppressive patients are prone to get sepsis, hypoimmune status was suggested to be the etiology of sepsis. However, the hypoimmune theory cannot successfully explain the proinflmmatory cytokines storm noted in sepsis. Both hyperimmune theory and hypoimmune theory have clinical and experimental evidences. However, they are contrary with each other. Here, I use the microarray study of whole blood of septic patients to propose a new theory: Sepsis is a syndrome of hyperactivity of innate immunity and hypoactivity of adaptive immunity. This new theory solves the above controversy.

## Material and Methods

### Microarray dataset

According to Dr. J. A. Howrylak's research in Physiol Genomics 2009, he collected total RNA from whole blood in sepsis and sepsis induced ARDS patients.(1) He tried to find out molecular signature of ARDS compared to sepsis patients. His dataset is available in Gene Expression Omnibus (GEO) www.ncbi.nlm.nih.gov/geo (assession number GSE 10474). I use his samples of sepsis patients from this dataset to do the further microarray analysis. The sample size is 21 patients with 35% mortality rate. The second dataset is from GSE20189 of Gene Expression Omnibus. This dataset was collected by Dr. Melissa Rotunno in Cancer Prevention Research 2011.(2) Molecular signature of early stage of lung adenocarcinoma was studied by microarray. I use the healthy control (sample size 21)whole blood RNA from this dataset to compare the septic patients. In this study, I perform further analysis to study peripheral leukocyte gene expression profiles of sepsis compared to those of healthy controls.

Statistical analysis

Affymetrix HG-U133A 2.0 genechip was used in both samples. RMA express software(UC Berkeley, Board Institute) is used to do normalization and to rule out the outliners of the above dataset. I rule out the potential outliners of samples due to the following criteria:

1. Remove samples which have strong deviation in NUSE plot
2. Remove samples which have broad spectrum in RLE value plot
3. Remove samples which have strong deviation in RLE-NUSE mutiplot
4. Remove samples which exceed 99% line in RLE-NUSE T2 plot

Then, Genespring XI software was done to analysis the significant expressed genes between ARDS and healthy control leukocytes. P value cut-off point is less than 0.05. Fold changecut-off point is >2.0 fold change. Benjamini-hochberg corrected false discovery rate was used during the analysis. Totally, a genelist of 3277 genes was generated from the HGU133A2.0 chip with 18400 transcripts including 14500 well-characterized human genes.

RT-PCR confirmation

Dr. J. A. Howrylak performed real time PCR for selected transcripts (cip1, kip2) by using TaqMan Gene Expression Assays (Applied Biosystems, Foster City, CA). In the second dataset, Dr. Melissa Rotunno also performed qRT-PCR test to validate the microarray results. RNA quantity and quality was determined by using RNA 600 LabChip-Aligent 2100 Bioanalyzer. RNA purification was done by the reagents from Qiagen Inc. All real-time PCRs were conducted by using an ABI Prism 7000 Sequence Detection System with the designed primers and probes for target genes and an internal control gene-GAPDH. This confirms that their microarray results are convincing compared to RT-PCR results.

## Results

RMA analysis of whole blood from healthy normal control

The RMA analysis was performed for RNA samples from whole blood of healthy control of the lung adenocarcinoma dataset. Raw boxplot, NUSE plot, RLE value plot, RLE-NUSE multiplot, and RLE-NUSE T2 plot were generated. Then, sample was included and excluded by using these graphs(Figure 1A, 1B, 1C, 1D, 1E). Because of the strong deviation in the T2 plot, the sample GSM506435 was removed for the further analysis.

RMA analysis of whole blood from septic patients

The RMA analysis was performed for RNA samples from whole blood of sepsis patients dataset. Raw boxplot, NUSE plot, RLE value plot, RLE-NUSE multiplot, and RLE-NUSE T2 plot were generated. Then, sample was included and excluded by using these graphs(Figure 2A, 2B, 2C, 2D, 2E). GSM265024 and GSM265030 are removed due to above criteria.

Toll-like signaling and heat shock protein expression in septic patients

According to the microarray analysis, Toll-like receptors 1, 2, 4, 5, 8 are up-regulated in sepsis.(Table 1) CD14 molecule and downstream signaling such as IRAK4 and TAB2 are also up-regulated. TLR1, 2, 4, 5, 8 are mediating anti-bacterial immune response. Thus, TH17-like proinflamatory cytokines such as IL-6 will be triggered. However, the negative TLR regulator-IRAK3 is 21 fold up-regulated. Thus, TLR 1, 2, 4, 5, 8 signaling may not successfully trigger proinflammatory cytokines. Other pathway such as CD14 may act as an important alternative pathway to trigger IL-6 and other TH17-like cytokines. Other pattern recognition receptors such as formyl peptide receptors (FPR)which can recognize specific bacterial antigen to trigger innate immunity are also differentially expressed. FPR1 is 7.6 fold down-regulated, but FPR2 is 4.7 fold up-regulated.

In table 2, we can see that many heat shock protein genes are up-regulated. Fever is a usual manifestation of sepsis. Thus, it is not surprising that heat shock proteins are expressed during sepsis. Among them, heat shock protein 70 (HSPA1A/1B) is 7 fold up-regulated. HSP70 can bind to TLR4 to trigger anti-bacterial TH17-like innate immunity. It is worth noting that HSP90AA1 is 13 fold down-regulated. HSP90 can bind to steroid receptor and prevent its action. If HSP90 is down-regulated, the action of steroid cannot be stopped. Thus, steroid related immune regulatory effect may be initiated during sepsis.

Antigen processing and antigen presentation genes in sepsis

In table 3 and table 4, we can see many cathepsin and proteasome genes are up-regulated. Up-regulated cathepsin genes include CTSA, CTSD, CTSC, CTSG, and CTSZ. But, CTSO and CTSW are down-regulated. Cathepsin W (CTSW) is related to CD8 T cell activation.(3) Up-regulated proteasome genes include PSMD13, PSMC6, PSMD12, PSMD5, PSMB6, and PSMD10. Down-regulated proteasome genes include PSMF1, PSMC2, and PSME4, Among them, PSMF1 is a proteasome inhibitor. Both cathepsins and proteasomes are important in the antigen processing pathways. We can see antigen processing after bacterial infection is intact.

In table 5, however, we can see all MHC related genes are down-regulated in leukocytes of septic patients. These down-regulated genes include HLA-DPB, HLA-DQA, HLA-DRB, HLA-DOB, HLA-DRA, HLA-DQB, Tapasin, MHC I related transcripts, HLA-B, and HLA-DPA. Among them, HLA-B is more than 11 fold down-regulated. MHC genes are keys to the antigen presentation to trigger adaptive immune reaction such as B cell or T cell activation. Since all the MHC related genes are down-regulated, antigen presentation during sepsis is likely to be impaired. This matches previous observations.(4)

TH17-like innate immune transcription factors in sepsis

In table 6, many immune related transcription factors are differentially regulated during sepsis. First of all, many innate immunity related transcription factors are up-regulated in septic patients. These include AP1(JunB and FosL2), NFIL3, ARNT, and CEBP(CEBPA, CEBPG, and CEBPD) genes. Aryl hydrocarbon receptor nuclear translocator(ARNT) plays an important role in the activation of TH17-like innate immunity. CEBP family genes are related to the activity of myeloid cells and granulocytes. CEBP genes are also related to the activation of acute response proteins. In addition, the inhibitor of NFkB, NFKBIA, is down-regulated in sepsis. It means that the activity of NFkB, an key innate immunity mediator, is up-regulated in septic patients. It is worth noting that two important transcription factors: High Mobility Group Box(HMGB) and Hypoxia inducible factor alpha(HIFα) are also up-regulated during sepsis. HMGB, a vital innate immunity mediator, is greater than nine fold up-regulation.

STAT1, a key transcription factor for TH1 and THαβ immunity, is down-regulated in sepsis. In addition, TBX21(T-bet), a key TH1 immune response driver, is also down-regulated. (Table 25) In addition, MafB which can suppress IFNαβ in THαβ immunity is up-regulated(5). Other TH2 related key transcription factors such as GATA3 and C-MAF are also down-regulated.(6) It means that TH1, TH2, and THαβ are down-regulated in sepsis. Surprisingly, key TH17 related transcription factors are also down-regulated including REL, STAT3, and RORA.(7) Besides, SOCS3, a negative regulator of the central TH17 transcription factor STAT3, is up-regulated. It means that TH17 helper cells cannot be successfully triggered. On the other hand, Treg and TGFβ signaling are up-regulated including STAT5B, IL-15, SMAD2, and SMAD4.(8, 9) TH17 and Treg associated aryl hydrocarbon receptor nuclear translocator(ARNT) is also up-regulated in sepsis.(10) Thus, Treg cells are likely to be activated in sepsis. This matches the previous observations that Treg cells are up-regulated during sepsis.

TH17-like and Treg related cytokines are up-regulated during sepsis

In table 7, many TH17-like and Treg related cytokines are up-regulated in septic patients. The whole TGFβ activation machinery is up-regulated including thrombospondin, CD36, and TGFβ itself. TGFA and IL-15 are also up-regulated. Besides, IL-6 is also up-regulated in sepsis. Thus, both key TH17 driven cytokines, TGFβ and IL-6, are activated in septic patients. However, full activation of TH17 helper cells also need a TCR signaling. IL-32, a TH1 related macrophage differentiation factor, is down-regulated.(11) TH22 mediators, IL1A is down-regulated and IL1RN (IL1 receptor antagonist is up-regulated. It means that TH22 is not activated in sepsis.

In table 8, cytokine receptors are differentially regulated in sepsis. On the contrary with cytokine, cytokine in a certain immunological pathway is usually down-regulated. Thus, since TH17-like immunity is activated. TGFBR3, IL6R, and IL17RA are all down-regulated. TGFβ receptor 3 is greater than 11 fold down-regulated, and interleukin 6 receptor is greater than 16 fold down-regulated. Treg is also triggered in sepsis, so TGFBR3, IL2RB, and IL7R are also down-regulated. TH1 related cytokine receptors, IFNGR1 and IFNGR2, are up-regulated. TH2 cytokine receptor, IL4R, is also up-regulated. As for THαβ immunity, IFNAR1 is up-regulated but IFNAR2 is down-regulated. TH22 cytokine receptors, IL1R1 and IL1R2, are up-regulated. Thus, TH1, TH2, THαβ, and TH22 are not activated during sepsis.

In table 9, important CSF receptors are up-regulated. These include CSF2 (GM-CSF) receptor α and β. GM-CSF can promote the proliferation of monocyte and granulocyte lineages. In table 10, many TNF related genes are differentially regulated. Up-regulated TNF related genes include TNFAIP6, TNFAIP8, TNFRSF1A and TNFSF10. Down-regulated TNF related genes include TNFRSF10C, TNFRSF9, and TNFSF14. TNFRSF1A is the major receptor of TNFα. Thus, both IL-1 receptor and TNFα receptor are up-regulated during sepsis. TNFSF10(TRAIL) is a pro-apoptotic factor, and TNFRSF10C is a receptor to prevent TRAIL induced apoptosis. Thus, TRAIL induced apoptosis pathway is activated in sepsis. TNFRSF9(4-1BB) and TNFSF14(CD258) are both important lymphocyte co-stimulatory molecules. Thus, lymphocyte costimulation is likely to be impaired at sepsis.

TH17 related chemokine up-regulation during sepsis

In table 11, we find out that TH17 related chemokine are up-regulated in septic patients. These chemokines include S100 binding proteins (S100A11, S100A8, S100A9, and S100P), CCR2(neutrophil chemokine receptor), hyaluronan-mediated motility receptor (HMMR), and chemokine-like factor(CKLF). THαβ related chemokine factors such as CX3CR1, XCL1, and XCL2 for NK cell recruitment are down-regulated. TH1 related chemokine factors such as CCL4 and CCR1 for macrophage/monocyte recruitment are also down-regulated. Besides, TH2 chemokine receptor CCR3 for eosinophil recruitment is also down-regulated. It is worth noting that CCR7, the chemokine receptor for central memory T cells, is greater than 5 fold down-regulated in sepsis. Thus, the generation of central memory T cell is likely to be impaired during sepsis.

In table 12, many prostaglandin and leukotriene genes are differentially regulated.

Prostaglandins and leukotrienes are important chemotaxis mediators. The key enzyme: leukotriene A4 hydrolase for synthesizing leukotriene B4, a potent PMN chemoattractant, is up-regulated. Besides, leukotriene B4 receptor is also up-regulated. Besides, the receptor of PGD2, a TH2 related effector molecule, is 10 fold down-regulated. Prostangin D synthetase is also down-regulated. In addition, the gene 15-hydroxyprostaglandin dehydrogenase (HPGD), which is responsible for shutting down prostaglandin, is 16 fold up-regulated. Key molecules including phospholipase A 2 and arachidonate 5-lipoxygenase to initiate leukotriene synthesis are also up-regulated in sepsis.

Th17-like innate immunity related effector molecule up-regulation in sepsis

In table 13, many acute response proteins are up-regulated. These acute phase proteins are up-regulated by IL-6 and CEBP proteins. These genes include amyloid proteins (APP and APLP2), pentraxin(PTX3), transferrin receptor(TFRC), CLEC (CLEC5A and CLEC1B), and defensins (DEFA1, DEFA1B, DEFA3, and DEFA4). These above proteins are innate immunity effector proteins to attack bacterial antigens non-specifically. Defensin A4 is greater than 6 fold up-regulated.

In table 14, the whole set complement machinery, an important effector component of innate immunity, is up-regulated. These include CD59, CD55, C1QB, ITGAM, CR1, CD46, C3AR1, ITGAX, C1QA, C1RL, C5AR1, and CD97. Thus, complement molecules are activated during sepsis. These complement molecules attack bacterial cell walls and membranes to cause their damage. However, complements may also cause harmful effect to the host.

In table 15, certain genes related to PMN phagocytosis and bacteria killing are up-regulated. Neutrophil cytosolic factor 1&4, the subunit of NADPH oxidase for ingested bacteria killing, are up-regulated in sepsis. Carboxypeptidase D (CPD), which can up-regulate nitric oxide, is also up-regulated during sepsis.(12) Nitric oxide is also a key effector molecule for ingested bacteria killing. CPD is greater than 6.9 fold up-regulated.

In table 16, PMN matrix metallopeptidases(MMP) and elastase are up-regulated. These protein enzymes can digest bacterial antigens as well as extracellular matrix. These genes include MMP8, MMP9, MMP25, and ELANE(elastase). In addition, tissue inhibitor of MMP, TIMP2, and serum inhibitors of elastase or proteinase, SERPINA1, SERPINB1, and SERPINB2, are also up-regulated. It means that PMN proteinases are dysregulated. It is worth noting that MMP8 is 32 fold up-regulated and MMP9 is 10 fold up-regulated.

In table 17, apoptosis machinery is up-regulated during sepsis. Up-regulated genes include casapase3, FAS, caspase5, program cell death 10(PDCD10), caspase 1, caspase4, and TRAIL. Down-regulated genes include CFLAR and FAIM3, both of which are apoptosis negative regulators. Thus, apoptosis is activated at sepsis. It matches previous observations that there is massive leukocyte-lymphocyte apoptosis during sepsis.

In table 18, many Fc receptor genes which mediate macrophage and neutrophil phagocytosis are up-regulated. These genes include IgG Fc receptor IIa(FCGR2A), IgE Fc receptor Ig(FCER1G), IgA Fc receptor(FCAR), IgG Fc receptor IIc(FCGR2C), IgG Fc receptor Ib, and IgG Fc receptor Ia/Ic(FCGR1A/1C). Besides, TH2 immunity related IgE Fc receptor Ia is 3.8 fold down-regulated. In addition, THαβ immunity related CD16 IgG Fc receptor expression is unchanged. TH17-related innate immunity is mediated by IgG(IgG2/IgG3) and IgA. Thus, TH17-like innate immunity with enhanced phagocytosis is noted during sepsis.

In table 19, many CD molecules are up-regulated or down-regulated during sepsis. These CD molecules are important immune response mediators. Among them, up-regulation of CD36, the thrombospodin receptor, means that TGFβ molecule is also up-regulated. Down-regulated CD2 molecule means that T cell activation pathway is impaired. In addition, down-regulation of CD40 means that activation of antigen presenting cells such as B cells is impaired. Besides, CD24 is usually down-regulated in memory B cells. However, during sepsis, CD24 is strongly up-regulated.

Coagulation , glycolysis, acidosis, and vasodilation gene dysregulation in sepsis

In table20, many coagulation related genes are dysregulated during sepsis. Actually, disseminated intracellular coagulopathy is a common manifestation of sepsis. Up-regulated coagulation genes include F13A1, F5, F8, GP1BB, PROS1, PLAUR, MCFD2, TFPI, F2RL1, ITGA2B, PDGFC, ITGB3, and THBD. Both coagulation factors and inhibitors are dysregulated in sepsis.

The whole glycolytic pathway enzymes are up-regulated during sepsis.(Table21) These include lactate dehydrogenase A, phosphoglycerate kinase 1, pyruvate kinase, 6-phosphofructo-2-kinase/fructose-2,6-biphosphatase 3, hexokinase 2, glycogen phosphorylase, 2,3-bisphosphoglycerate mutase, hexokinase 3, glucose-6-phosphate isomerase, 6-phosphofructo-2-kinase/fructose-2,6-biphosphatase 2, glyceraldehyde-3-phosphate dehydrogenase, enolase 1, and phosphoglycerate kinase 1. In addition, the enzyme, pyruvate dehydrogenase kinase, which can stop pyruvate to form acetyl-CoA is up-regulated. The enzyme, pyruvate dehyrogenase phosphatase, which can facilitate pyruvate to form acetyl-CoA to enter aerobic citric acid cycle is down-regulated in sepsis. Thus, pyruvate can keep on forming lactate in anaerobic pathway during sepsis.

Concurrently, $H^+$-ATPases are also up-regulated during sepsis(Table22). In my previous article(paper in press), I find out the coupling between glycolytic enzymes and $H^+$-ATPases during falciparum malarial infection. Here, I also find out up-regulated $H^+$-ATPases including ATP6V0B, ATP6V0E1, ATP6AP2, ATP6V1C1, TCIRG1, ATP6V1D, ATP11B, and ATP11A. Besides, carbonic anhydrase IV & II, which can produce $H_2CO_3$, are up-regulated in sepsis. Thus, this can help to explain the acidosis during sepsis.

Hypotension is a complication of sepsis. Septic shock is related to global vasodilation. In table 23, several key vasodilation genes are up-regulated. These genes including angiotensinase c (PRCP), adrenomedullin(ADM), and monoamine oxidase A(MAO-A). Angiotensin is a very strong vasopressor, and angiotensinase can metabolize angiotensin to inhibit its function. Adrenomedullin is a potent vasodilator. MAO-A can metabolize three key endogenous vasoconstrictors: dopamine, epinephrine, and norepinephrine. Thus, elevated MAO-A can prevent the effect of hypertensive agents. Thus, the findings can help to explain the etiology of septic shock.

Failure of NK, B, T lymphocyte adaptive immunity during sepsis

Lymphocytes play important roles in adaptive immunity. In sepsis, the three major lymphocyte populations: NK cells, T cells, and B cells are all down-regulated. Thus, lymphocyte adaptive immunity fails to induce during sepsis. This is very important is sepsis pathogenesis.

In table 24, majority of NK cell related genes are down-regulated including NK cell killer receptors (NKTR, KLRK1, LAIR2, KLRD1, KLRG1, KLRB1, and KLRF1), granzymes (GZMA, GZMK, GZMB, and GZMH), and perforin (PRF1). Thus, NK lymphocytes are inactivated or even suppressed during sepsis. NK cells are very important in THαβ immunity, and it also implies that THαβ immunity is inhibited during sepsis.

In table 25, many T cell related genes are also down-regulated. These down-regulated genes include TCR genes(TRAC, TARP, TRBC1/C2, TRD@, TRGC2, and TRDV3), CD costimulatory molecules(CD3E, CD8A, CD3G, LY9, CD3D,CD2), T cell specific transcription factors(IKZF1, TCF7, NFAT5, NFATC3, TCF7L2, NFATC2IP, TBX21, ID2, and ID2B), granzyme/perforin (GZMA, GNLY, GZMK, GZMB, GZMH, and PRF1), and TCR downstream signaling(ZAP70 and LCK)(13). Thus, the whole-set of T cell activation machinery is suppressed. Both CD4 helper T cells and CD8 cytotoxic T cells are inactivated and down-regulated in septic patients.

In table 25, B cell genes are differentially regulated. Up-regulated B cell related genes include immunoglobulin light chain(IGK, IGJ, and IGKV1-5), immunoglobulin heavy chain(IGHG1/G2 and IGHA1/A2), and B cell suppressive transcription factors(BCL6 and IBTK)(14). Down-regulated genes include B cell stimulatory transcription factor (PAX5 and IKZF1), immunoglobulin heavy chain(IGHM), BCR signaling(FYN and LYN), and PI3K signaling(PIK3CB, PIK3IP1, PIK3CG, and PIK3R1)(15-17). The negative regulator of PI3K signaling, PTEN, is 4.6 fold up-regulated. BCL6 and IBTK can inhibit B cell differentiation and activation. PI3K signaling is the downstream stimulatory pathway of B cell activation. Thus, BCR signaling appears to be suppressed during sepsis. Although Ig light chain and IgA/IgG2 are up-regulated, this may be due to the isotype switch effect of up-regulated immunosuppressant TGF-β. The key immunoglobulin, IgM, for bacteria defense is down-regulated. It implies that B cell adaptive immunity is also impaired at sepsis.

## Discussion

Despite of current antibiotics treatment, sepsis still causes a very high mortality. The pathophysiology of sepsis is still unclear.(18, 19) The most dominant theory for sepsis mechanism is hyperimmune.(20) Hyperimmunity with cytokine storm was observed in sepsis by Dr. Lewis Thomas.(21) He suggested the symptom and sign in sepsis is due to the overactivity of pro-inflammatory cytokines in a NEJM paper. His theory is widely accepted. Based on his hyperimmune theory, many therapeutic strategies were developed. Most famous approach is the anti-TNF agent in sepsis clinical trials. Because the pro-inflammatory cytokine TNFα is up-regulated in sepsis, use of anti-TNF agent should help to control sepsis. However, the result is opposite. Usage of anti-TNF agent increase the sepsis mortality rate.(22-24) Thus, the sepsis-hyperimmune theory is doubtful.

The other sepsis pathophysiology theory emerged. This is the hypoimmune theory. Because immunocompromised patients are prone to develop sepsis, hypoimmune should be related to the cause of sepsis.(25) In addition, massive effector lymphocyte apoptosis, depletion of dendritic cells, and elevated Treg cells are noted during sepsis.(26-30) In previous reports, down-regulation of co-stimulatory molecules and MHC are noted in septic patients.(31) In addition, B cells play important roles in recovery from sepsis status.(32) This hypothesis is not accepted by most scientists because it cannot explain the observed cytokine storm during sepsis. Thus, both theories have some evidence support and both are only partially correct.

Thus, the third theory proposed. This is the sequence theory. There is hyperimmune first during sepsis, and then hypoimmune follows. This theory tried to incoperate both theories. However, it is not clear why there will be such sequential immune response. There is no existing immunological mechanism to explain this sequential effect. Why does the hyperimmune happen first? Why does the hyperimmune become to be hypoimmune? In addition, immunodeficiency patients are easily get sepsis. Then, why do these immunodeficiency patients easily develop hyperimmune status first? Current sepsis theory cannot well explain this.

In this study, I use microarray analysis to demonstrate that sepsis is actually a hyperactivity of innate immunity and hypoactivity of adaptive immunity. Thus, it can well explain the co-existence of hyperimmune and hypoimmune. The hypoimmune adaptive immunity explains why immunocompromised patients tend to suffer from sepsis easily. The hyperimmune innate immunity explains why pro-inflammatory cytokine storm is observed at sepsis. The adaptive immune dysfunction with lack of T helper cells is the key to sepsis pathogenesis. Thus, block TH17 related cytokines such as TNFα can further stop the successful generation of TH17 helper cells to initiate adaptive immunity to combat extracellular bacteria. Thus, it can explain why TNF blockade increase the mortality rate of sepsis patients.

In the microarray study, I find out evidences to support my theory. The whole blood from septic patients can reflect the leukocyte expression patterns. I find out that innate immunity related genes are significantly up-regulated. These genes include CD14, TLR1,2,4,5,8, HSP70, CEBP proteins, AP1(JUNB, FOSL2), TGF-β, IL-6, TGF-α, CSF2 receptor, TNFRSF1A, S100A binding proteins, CCR2, formyl peptide receptor2, amyloid proteins, pentraxin, defensins, CLEC5A, whole complement machinery, CPD, NCF, MMP, neutrophil elastase, caspases, IgG and IgA Fc receptors(CD64, CD32), ALOX5, PTGS, LTB4R, LTA4H, and ICAM1. I also find out that majority of adaptive immunity genes are down-regulated including MHC related genes, TCR genes, granzymes/perforin, CD40, CD8, CD3, TCR signaling, BCR signaling, T & B cell specific transcription factors, NK killer receptors, and TH17 helper specific transcription factors(STAT3, RORA, REL). In addition, Treg related genes are up-regulated including TGFβ, IL-15, STAT5B, SMAD2/4, CD36, and thrombospondin. Up-regulated regulatory cells during sepsis are also shown in other previous studies. Up-regulated Treg related genes can also suppress the adaptive immunity in sepsis. These all support my sepsis pathogenesis.

Sepsis is also related to several complications such as disseminated intravascular coagulation(DIC), hypotension/shock, and lactate acidosis.(33) In this microarray analysis, I find out that many coagulation related genes are up-regulated during sepsis including factor5, factor8, facto13, protein S, plasminogen receptor, ITGA2B, ITGB3, and thrombomodulin. Thus, it can help to explain the mechanism of DIC during sepsis. Several hypotensive agents are also up-regulated in sepsis including angiotesinase C, adrenomodullin, and MAO-A. This can help to explain septic shock pathophysiology. The whole set of glycolytic enzymes are up-regulated during sepsis including LDHA, PGK1, PKM2, PFKFB3, HK2, PYGL, BPGM, HK3, PDK3, GPI, PFKFB2, GAPDH, and ENO1. In addition, glycolytic enzyme coupled $H^+$-ATPase genes are also up-regulated. These can explain the lactate acidosis noted during sepsis.

Bacteria have strategies to suppress host immunity for their survival, especially the adaptive immunity.(34) In conclusion, after knowing the pathogenesis of sepsis, we can develop better preventive and therapeutic agents to control sepsis. The impairment of adaptive immunity could be more important than the overactivation of innate immunity during sepsis. Thus, we may use medications to activate host adaptive immunity such as T helper cells to combat sepsis. In addition, we can also develop therapeutic strategies to cope with sepsis related complications such as DIC, hypotension, and lactate acidosis. Hopefully, the detrimental illness-sepsis will be overcome one day.

**Author's information**

Wan-Chung Hu is a MD from College of Medicine of National Taiwan University and a PhD from vaccine science track of Department of International Health of Johns Hopkins University School of Public Health. He is a postdoctorate in Genomics Research Center of Academia Sinica, Taiwan. His previous work on immunology and functional genomic studies were published at *Infection and Immunity* 2006, 74(10):5561, *Viral Immunology* 2012, 25(4):277, and *Malaria Journal* 2013,12:392. He proposed THαβ immune response as the host immune response against viruses.

Figure legends

Figure 1. RMA express plot for selecting samples in normal healthy controls.

1-A NUSE boxplot for normal control

1-B RLE boxplot for normal control

1-C RLE-NUSE multiplot for normal control

1-D RLE-NUSE T2 plot for normal control

1-E Raw data Boxpolt for normal control

Figure 2. RMA express plot for selecting samples in septic patients.

2-A NUSE boxplot for septic patients

2-B RLE boxplot for septic patients

2-C RLE-NUSE multiplot for septic patients

2-D RLE-NUSE T2 plot for septic patients

2-E Raw data Boxplot for septic patients

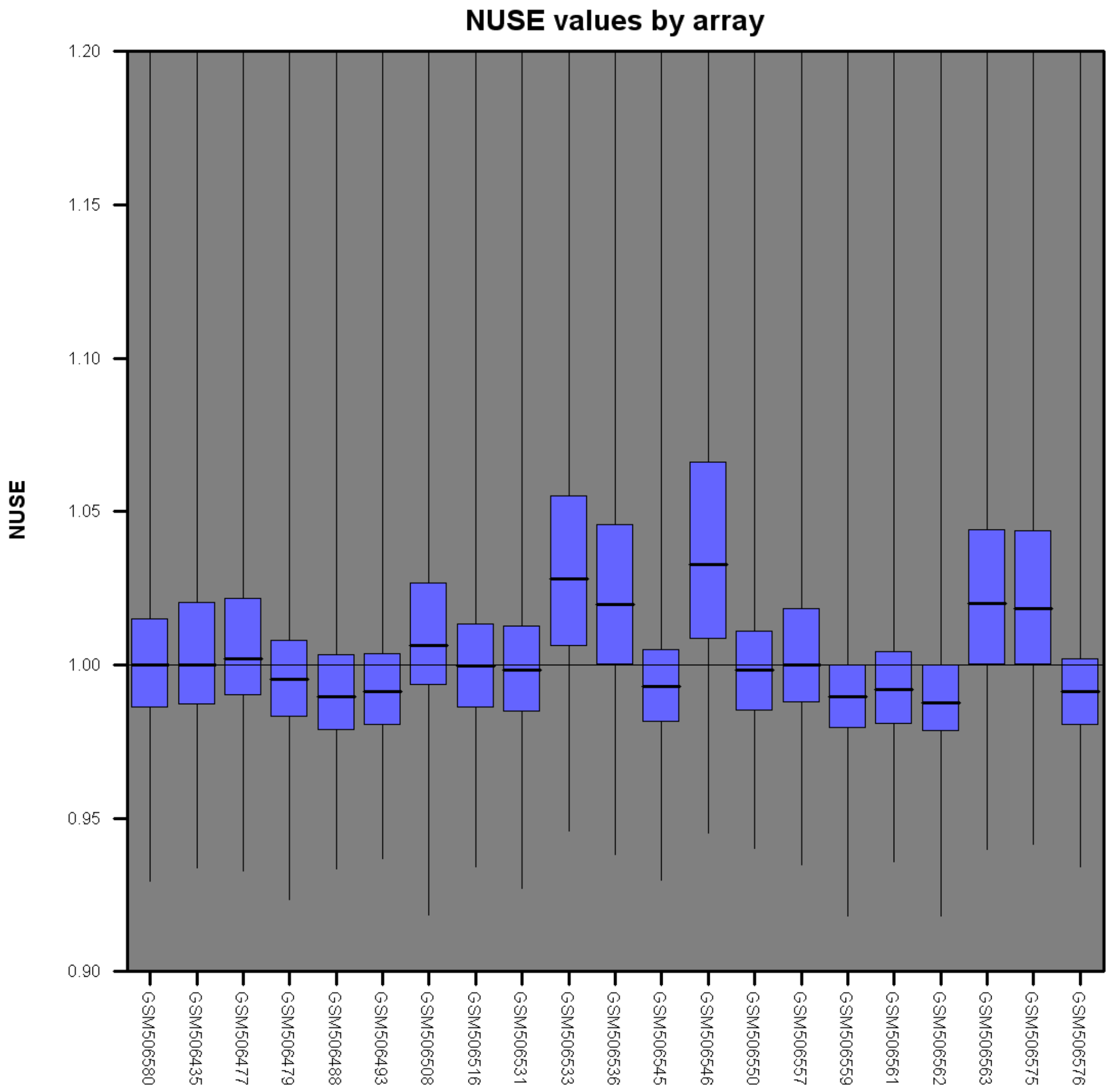

**Figure 1A**

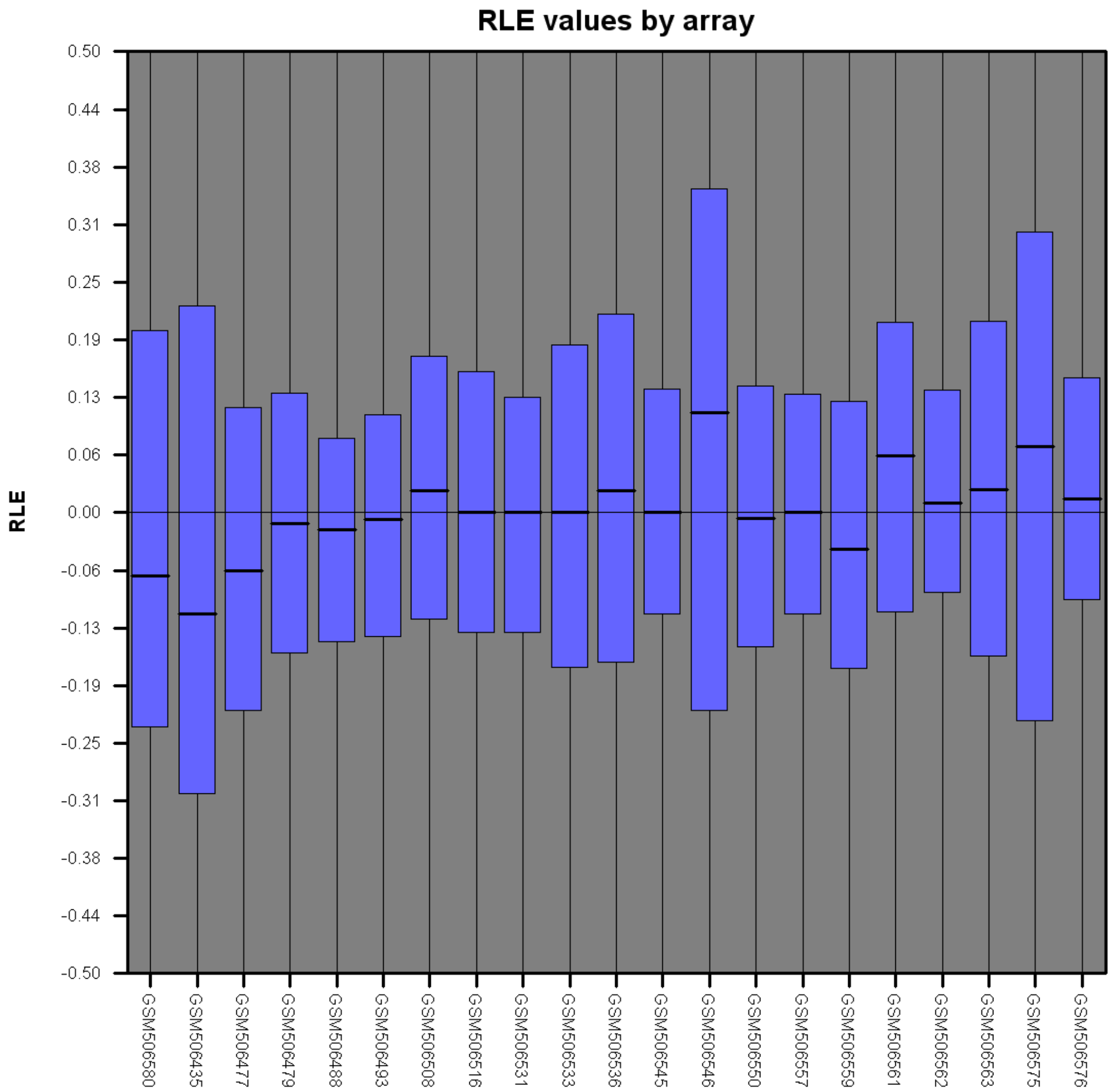

**Figure 1-B**

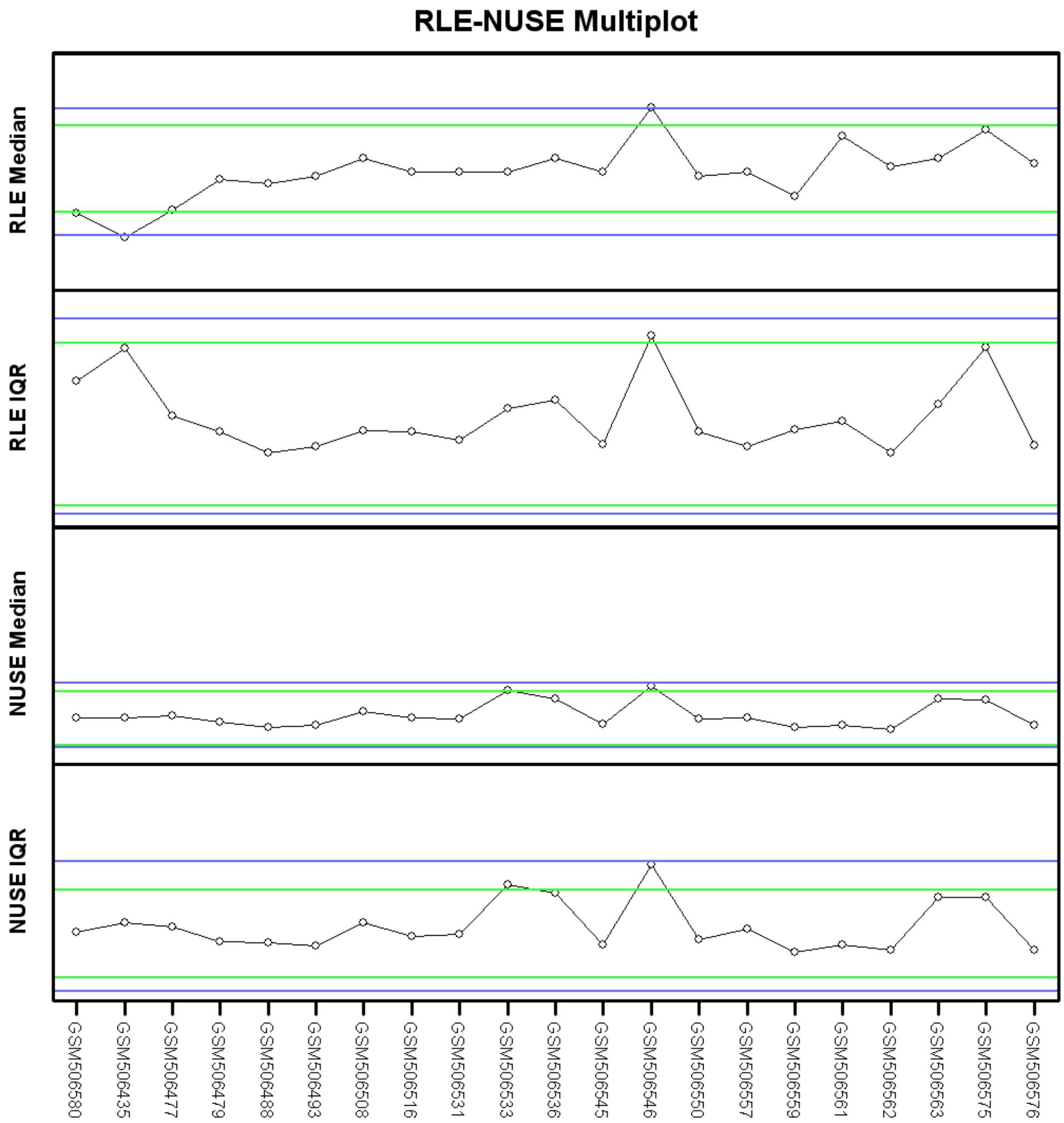

**Figure 1-C**

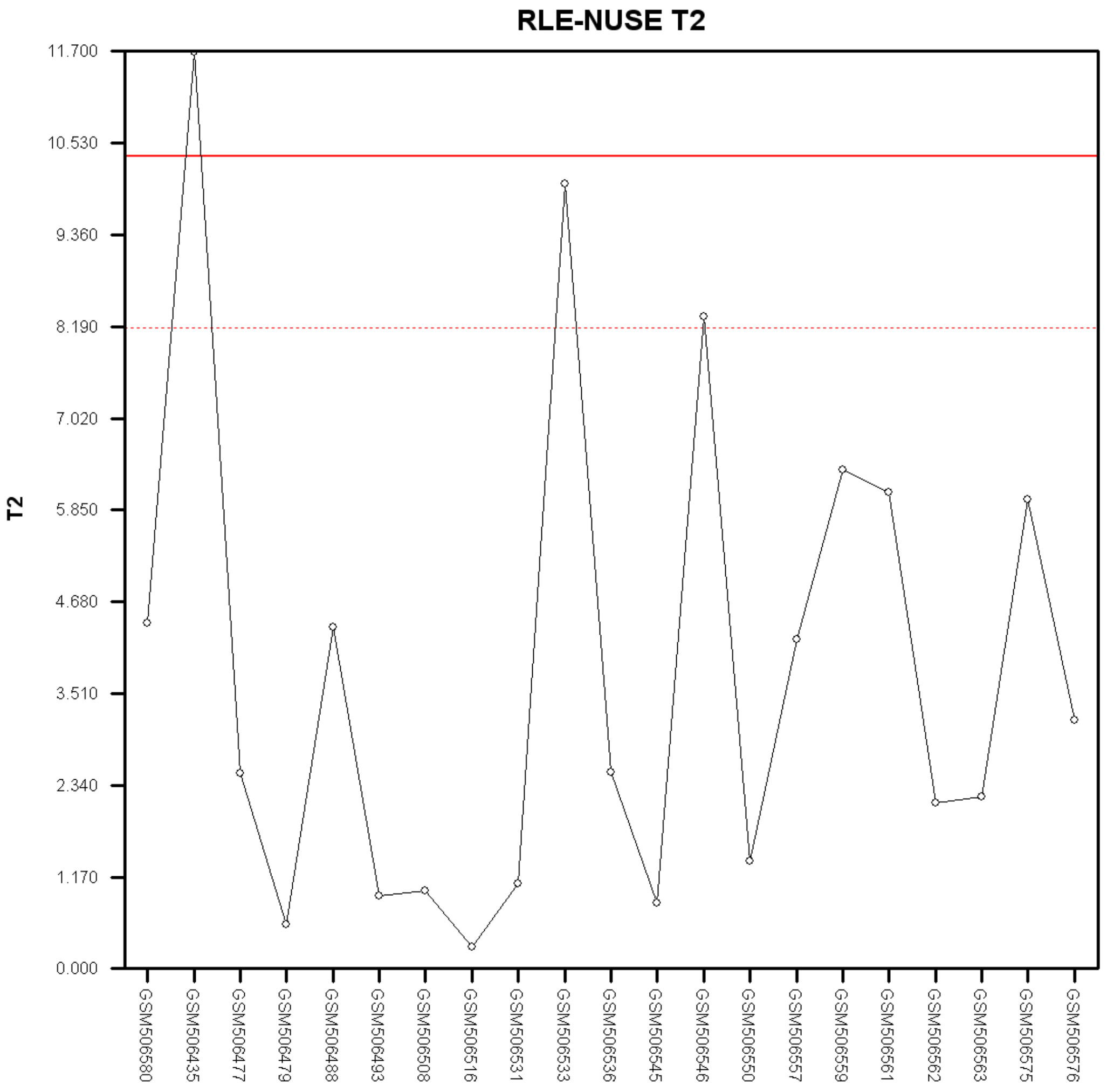

**Figure 1-D**

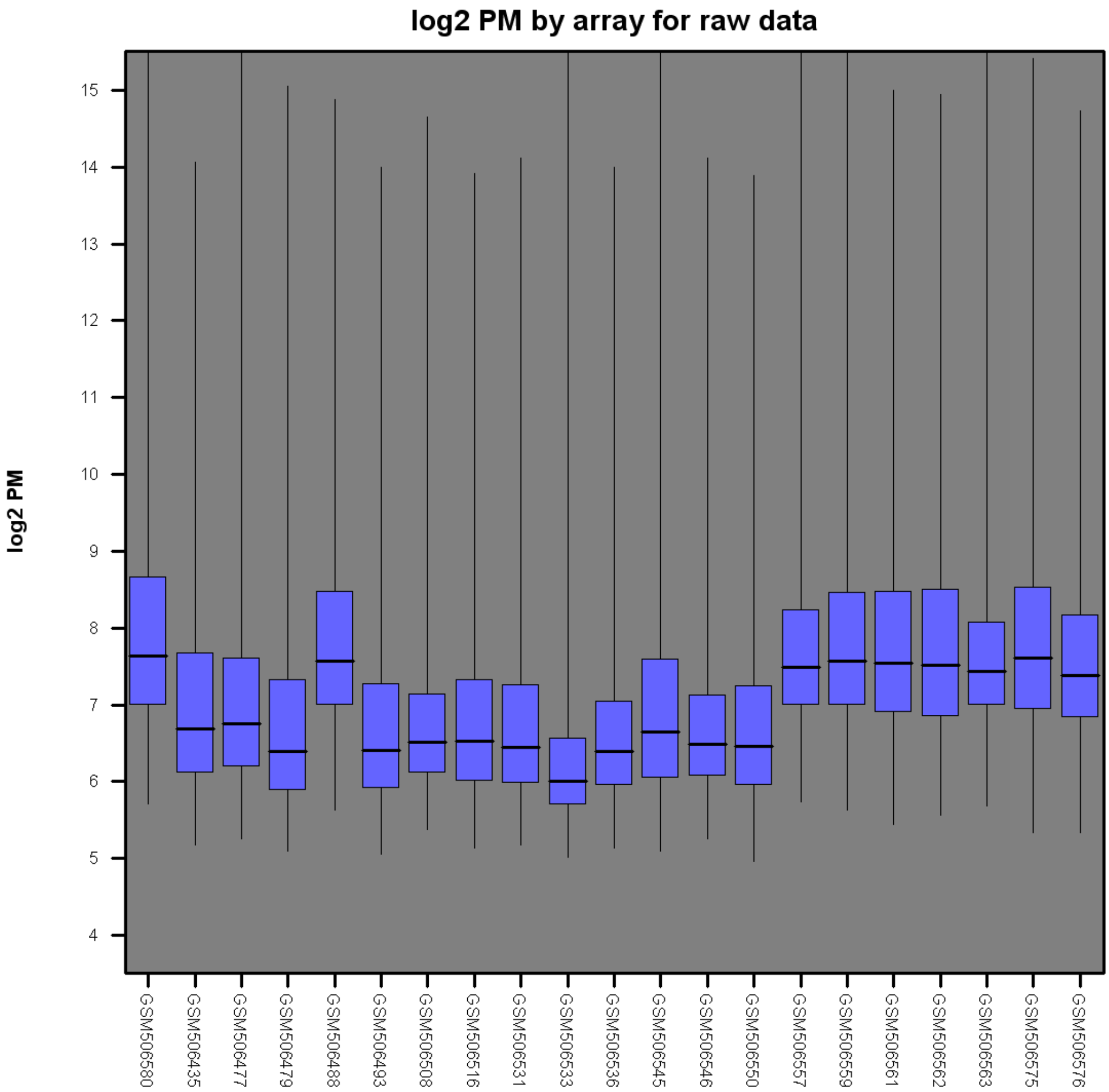

**Figure 1-E**

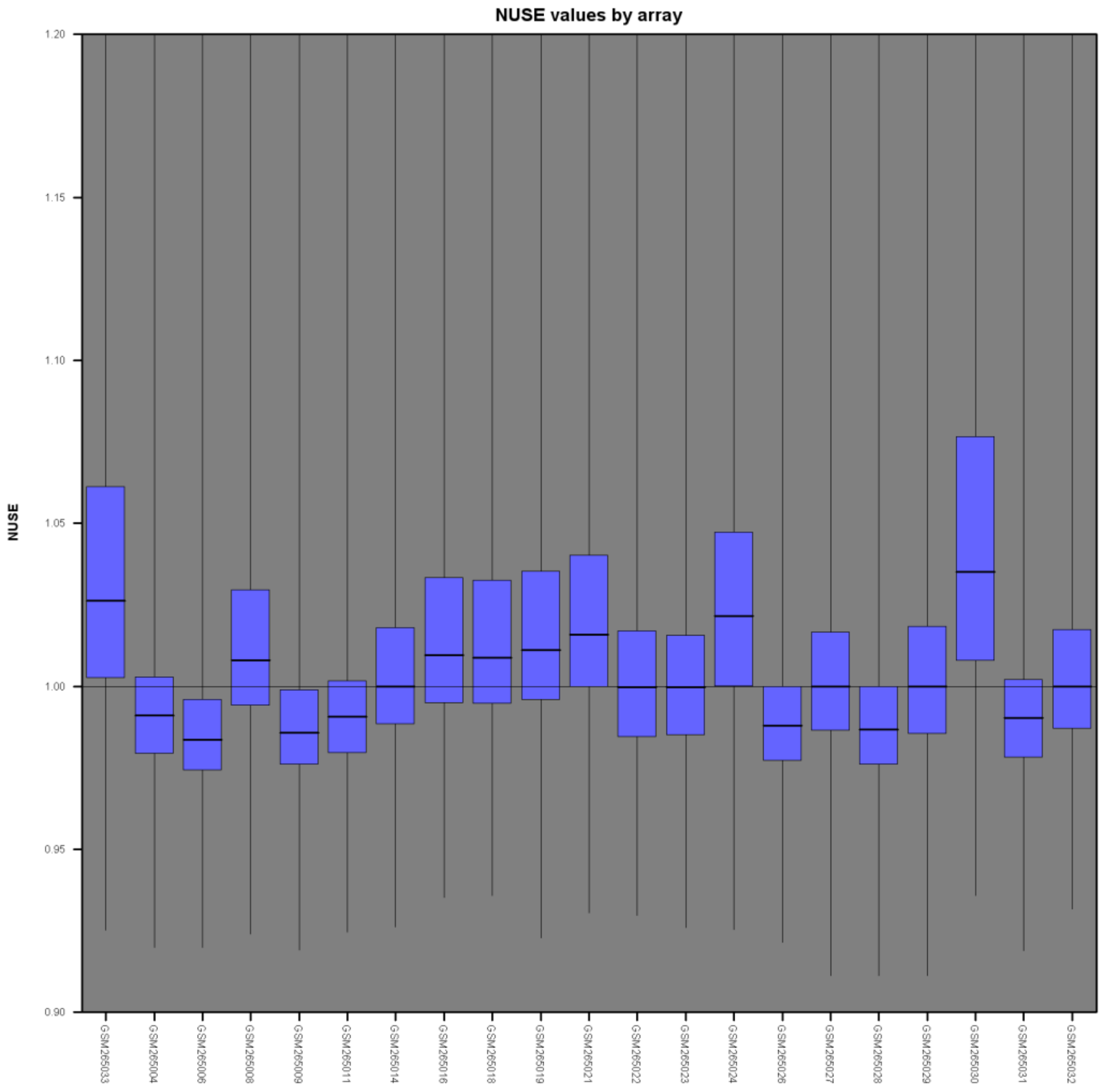

**Figure 2-A**

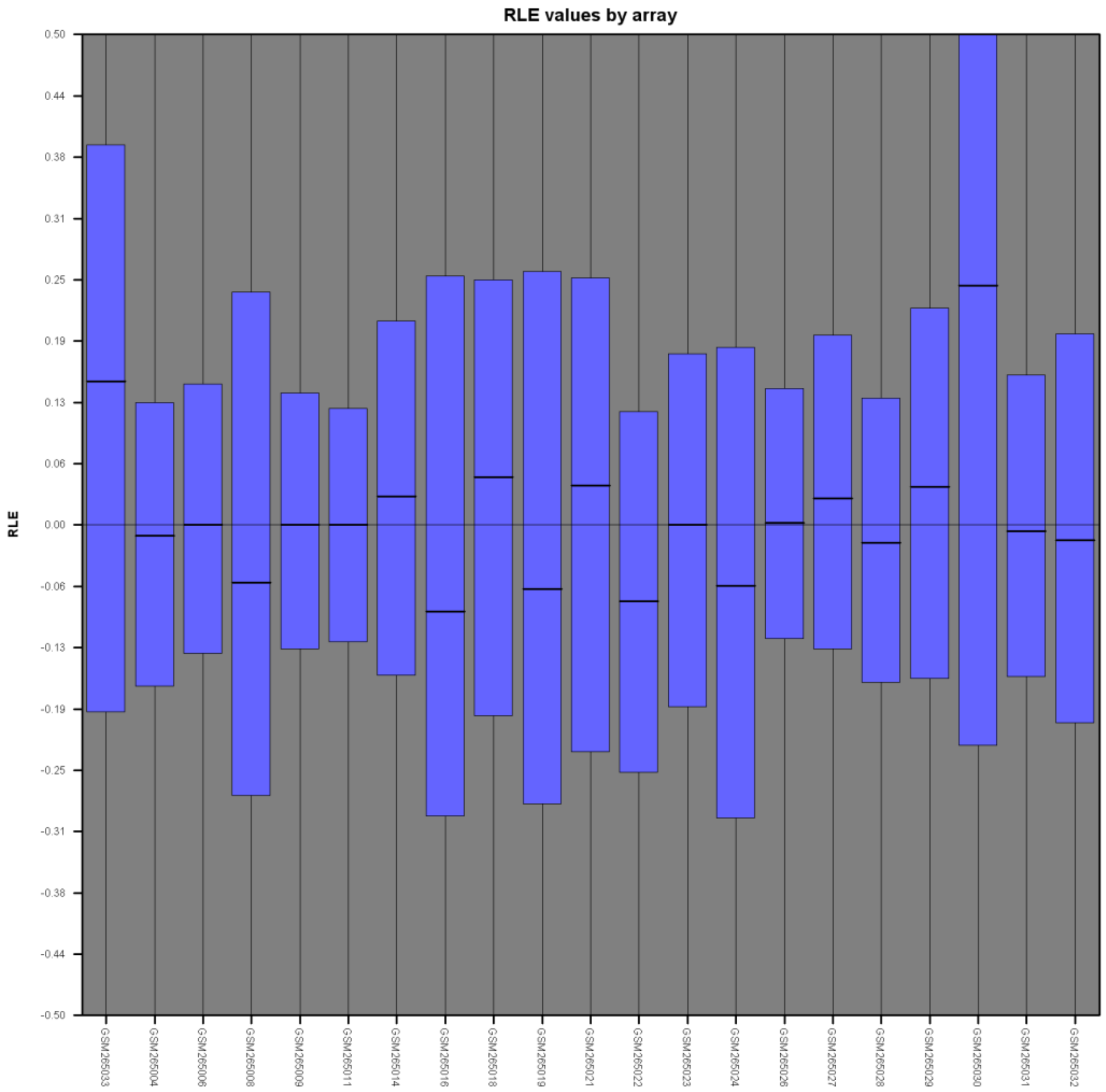

**Figure 2-B**

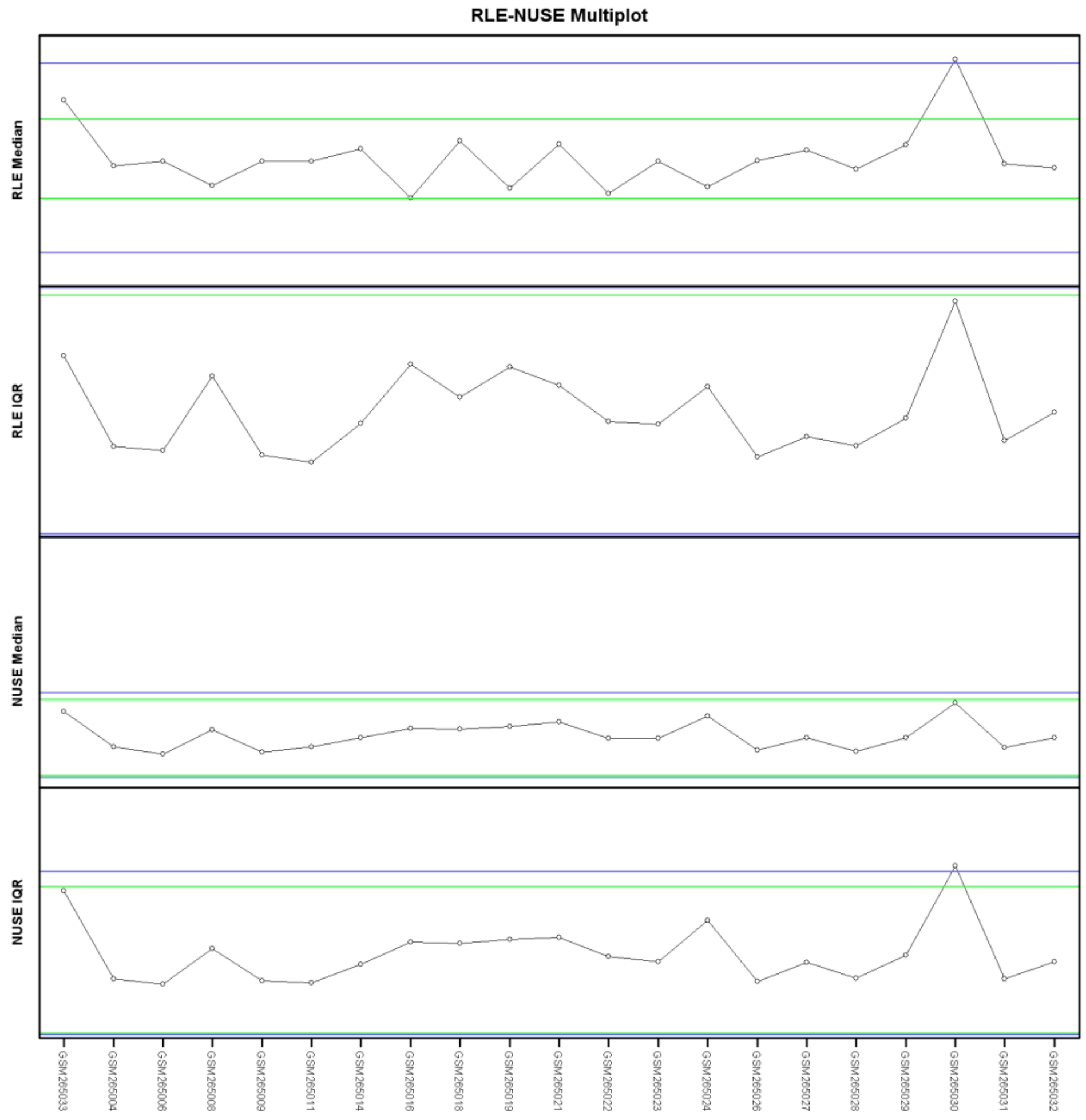

**Figure 2-C**

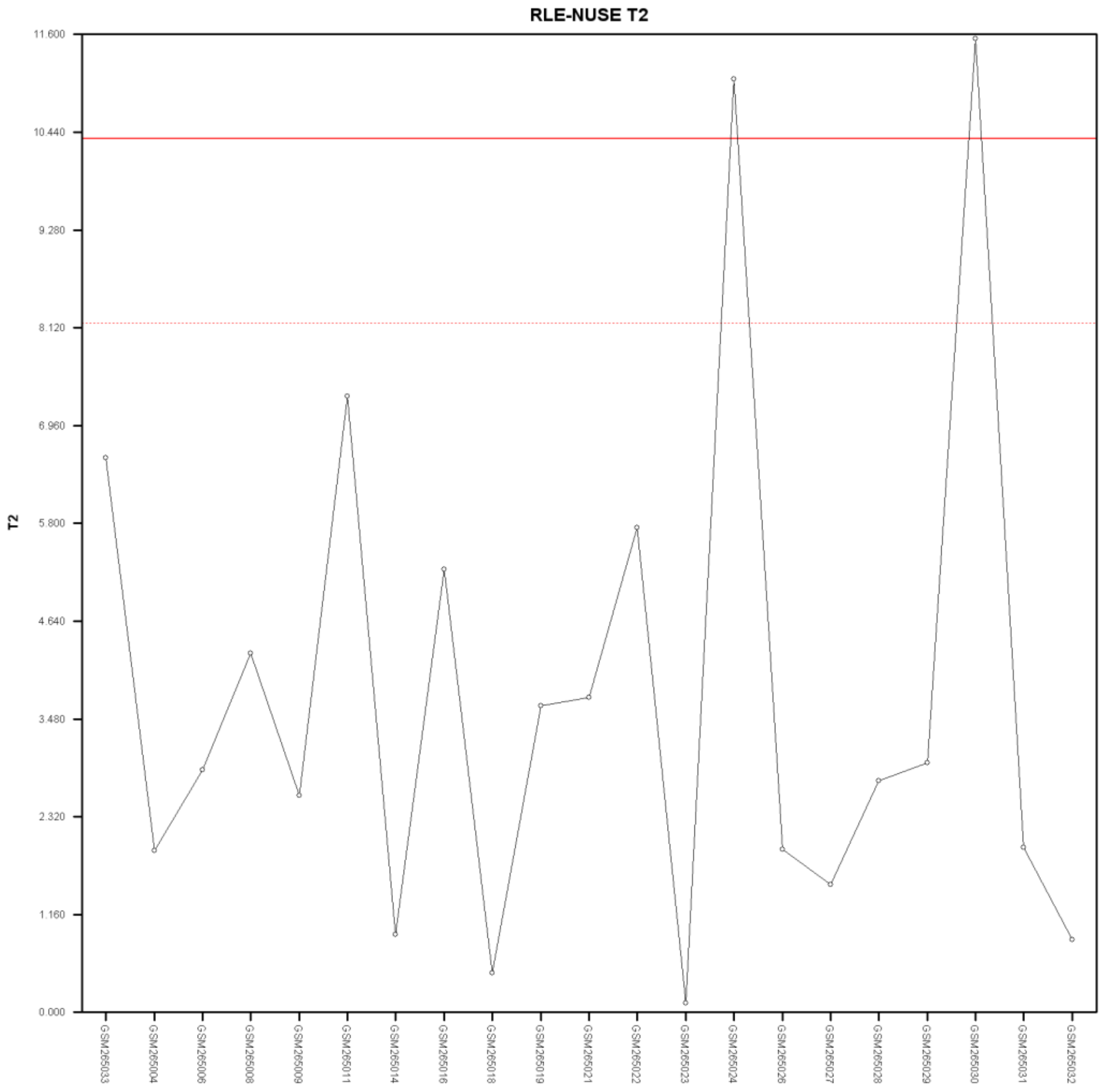

**Figure 2-D**

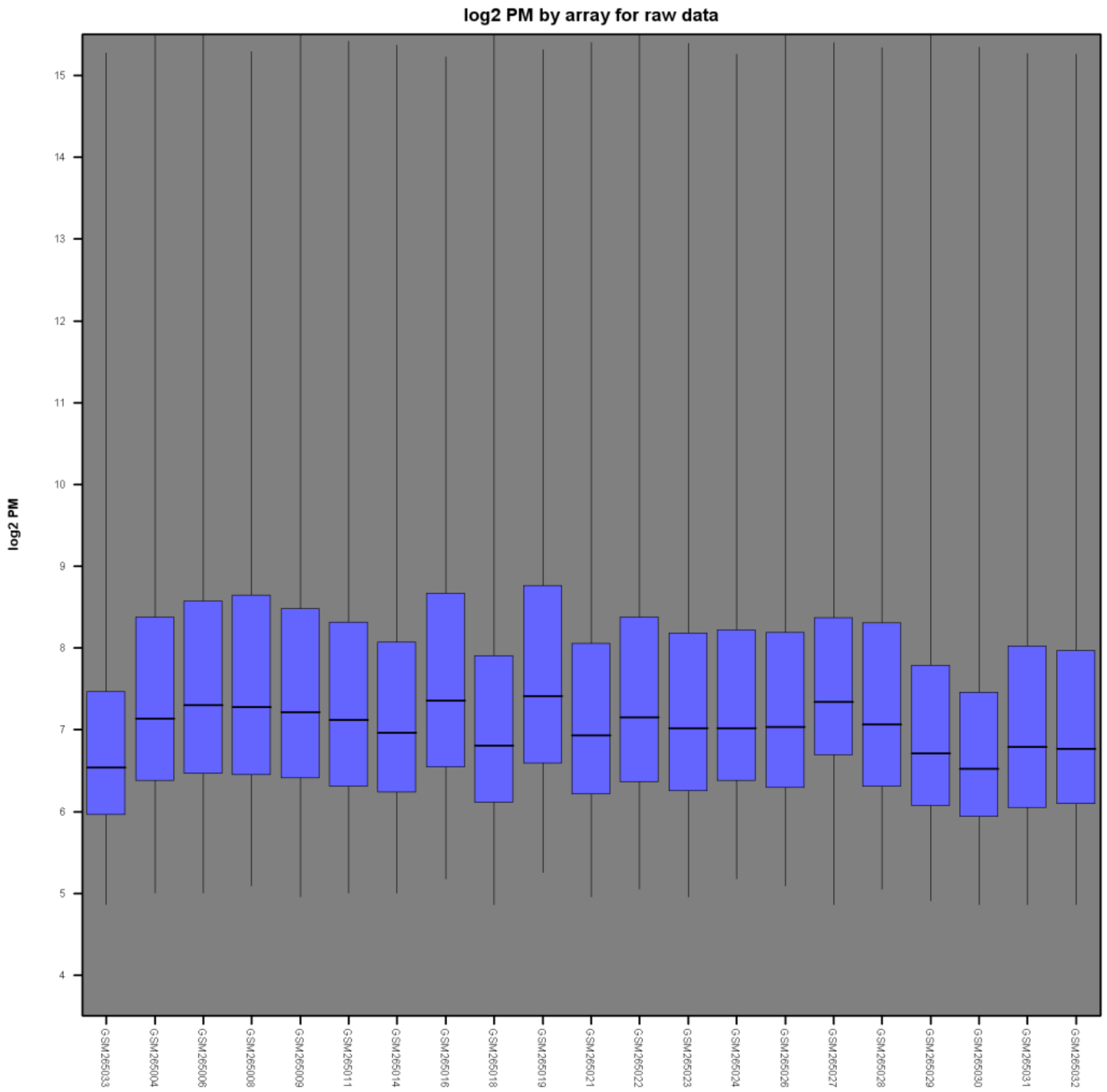

**Figure 2-E**

Table 1. TLR

| Probe ID | Pvalue | Arrow | Fold | Gene |
|---|---|---|---|---|
| 201743_at | 1.37E-04 | up | 2.176301 | CD14 |
| 204924_at | 1.45E-10 | up | 3.380896 | TLR2 |
| 210166_at | 9.16E-08 | up | 2.395743 | TLR5 |
| 210176_at | 0.001131 | up | 2.073326 | TLR1 |
| 213817_at | 3.14E-13 | up | 21.0364 | IRAK3 |
| 219618_at | 1.89E-09 | up | 2.692996 | IRAK4 |
| 220832_at | 4.76E-09 | up | 5.164016 | TLR8 |
| 221060_s_at | 6.62E-07 | up | 3.331681 | TLR4 |
| 212184_s_at | 2.03E-05 | up | 2.61743 | TAB2 |
| 221705_s_at | 8.46E-10 | down | 2.07506 | SIKE1 |
| 205118_at | 1.05E-10 | down | 7.612908 | FPR1 |
| 210772_at | 2.06E-08 | up | 4.776773 | FPR2 |
| 210773_s_at | 2.95E-06 | up | 4.516464 | FPR2 |

Table 2. HSP

| Probe ID | Pvalue | Arrow | Fold | Gene |
|---|---|---|---|---|
| 200598_s_at | 4.02E-04 | down | 2.112279 | HSP90B1 |
| 200599_s_at | 6.27E-04 | up | 2.143909 | HSP90B1 |
| 200800_s_at | 8.98E-09 | up | 2.664816 | HSPA1A /1B |
| 200941_at | 1.29E-10 | up | 2.175182 | HSBP1 |
| 200942_s_at | 7.06E-08 | up | 2.197023 | HSBP1 |
| 202557_at | 1.10E-05 | up | 2.82175 | HSPA13 |
| 202558_s_at | 7.12E-05 | up | 2.112017 | HSPA13 |
| 202581_at | 1.98E-14 | up | 7.20045 | HSPA1A/1B |
| 202842_s_at | 6.81E-06 | up | 2.73954 | DNAJB9 |
| 202843_at | 8.03E-07 | up | 2.087716 | DNAJB9 |
| 206782_s_at | 1.21E-09 | up | 2.374709 | DNAJC4 |
| 208810_at | 3.57E-04 | up | 2.196654 | DNAJB6 |
| 209015_s_at | 9.66E-09 | up | 2.548422 | DNAJB6 |
| 209157_at | 1.97E-10 | up | 2.893105 | DNAJA2 |
| 210338_s_at | 1.85E-05 | down | 2.06926 | HSPA8 |
| 211936_at | 1.05E-07 | up | 2.110743 | HSPA5 |
| 211969_at | 3.39E-17 | down | 13.51482 | HSP90AA1 |
| 212467_at | 1.01E-07 | up | 3.702917 | DNAJC13 |
| 212911_at | 1.88E-13 | up | 3.009307 | DNAJC16 |
| 219237_s_at | 2.18E-04 | down | 2.275608 | DNAJB14 |

Table 3. Cathepsin

| Probe ID | Pvalue | Arrow | Fold | Gene |
|---|---|---|---|---|
| 200661_at | 6.28E-10 | up | 2.550876 | CTSA |
| 200766_at | 3.53E-12 | up | 3.746886 | CTSD |
| 201487_at | 7.15E-06 | up | 2.584763 | CTSC |
| 203758_at | 4.23E-08 | down | 2.3372 | CTSO |
| 205653_at | 1.36E-04 | up | 3.234722 | CTSG |
| 210042_s_at | 2.94E-07 | up | 3.044302 | CTSZ |
| 214450_at | 2.16E-06 | down | 2.202781 | CTSW |

Table 4. Proteasome

| Probe ID | Pvalue | Arrow | Fold | Gene |
|---|---|---|---|---|
| 201052_s_at | 8.16E-06 | down | 2.218188 | PSMF1 |
| 201067_at | 3.68E-20 | down | 4.692738 | PSMC2 |
| 201232_s_at | 5.58E-11 | up | 2.577756 | PSMD13 |
| 201699_at | 2.88E-08 | up | 4.61153 | PSMC6 |
| 202352_s_at | 1.45E-08 | up | 2.49783 | PSMD12 |
| 202353_s_at | 1.04E-07 | up | 2.4536 | PSMD12 |
| 202753_at | 6.55E-08 | up | 2.001992 | PSMD6 |
| 203447_at | 1.23E-10 | up | 2.450336 | PSMD5 |
| 208805_at | 6.04E-05 | up | 2.094127 | PSMA6 |
| 208827_at | 4.41E-09 | up | 2.513269 | PSMB6 |
| 212220_at | 1.31E-09 | down | 3.309962 | PSME4 |
| 219485_s_at | 1.85E-07 | up | 2.271658 | PSMD10 |

Table 5. MHC

| Probe ID | Pvalue | Arrow | Fold | Gene |
|---|---|---|---|---|
| 201137_s_at | 5.80E-04 | down | 2.085101 | HLA-DPB1 |
| 203290_at | 2.56E-08 | down | 5.193201 | HLA-DQA1 |
| 204670_x_at | 6.77E-08 | down | 2.848446 | HLA-DRB1/B4 |
| 205671_s_at | 1.27E-04 | down | 2.016965 | HLA-DOB |
| 208306_x_at | 1.53E-06 | down | 2.437791 | HLA-DRB1 |
| 208894_at | 8.06E-07 | down | 2.762713 | HLA-DRA |
| 209312_x_at | 1.24E-06 | down | 2.674953 | HLA-DRB1/B4/B5 |
| 209823_x_at | 8.65E-04 | down | 2.083243 | HLA-DQB1 |
| 210294_at | 7.08E-10 | down | 2.247553 | TAPBP |
| 210528_at | 1.28E-05 | down | 2.565486 | MR1 |
| 210982_s_at | 4.46E-05 | down | 2.140575 | HLA-DRA |
| 211944_at | 5.60E-22 | down | 7.377162 | BAT2L2 |
| 211947_s_at | 9.14E-14 | down | 4.692785 | BAT2L2 |
| 211948_x_at | 3.66E-28 | down | 11.74736 | BAT2L2 |
| 211990_at | 5.10E-06 | down | 3.188215 | HLA-DPA1 |
| 211991_s_at | 1.47E-05 | down | 2.428313 | HLA-DPA1 |
| 212384_at | 8.83E-15 | down | 2.983359 | HLABAT1 |
| 212671_s_at | 0.002545 | down | 2.265915 | HLA-DQA1/A2 |
| 213537_at | 7.83E-05 | down | 2.338689 | HLA-DPA1 |
| 214052_x_at | 4.45E-14 | down | 2.404988 | BAT2L2 |
| 214055_x_at | 1.16E-24 | down | 9.415468 | BAT2L2 |
| 215193_x_at | 2.90E-06 | down | 2.522364 | HLA-DRB1/B3/B4 |
| 221491_x_at | 1.50E-06 | down | 2.247077 | HLA-DRB1/B3/B4/B5 |

Table 6. Transcription factor

| Probe ID | Pvalue | Arrow | Fold | Gene |
|---|---|---|---|---|
| M97935_MA_at | 1.92E-04 | down | 2.010955 | STAT1 |
| 201473_at | 3.65E-09 | up | 2.459731 | JUNB |
| 201502_s_at | 9.16E-07 | down | 2.364327 | NFKBIA |
| 202527_s_at | 5.77E-09 | up | 3.237444 | SMAD4 |
| 203075_at | 3.46E-06 | up | 2.133481 | SMAD2 |
| 203077_s_at | 4.90E-07 | up | 2.370214 | SMAD2 |
| 203574_at | 4.37E-10 | up | 5.176725 | NFIL3 |
| 204039_at | 4.62E-08 | up | 2.059098 | CEBPA |
| 204203_at | 9.92E-07 | up | 2.172468 | CEBPG |
| 205026_at | 1.66E-09 | up | 2.20184 | STAT5B |
| 205841_at | 1.02E-13 | up | 4.655113 | JAK2 |
| 205842_s_at | 6.01E-07 | up | 2.878639 | JAK2 |
| 206035_at | 7.43E-10 | down | 2.106359 | REL |
| 206036_s_at | 8.46E-12 | down | 4.746634 | REL |
| 206359_at | 3.22E-07 | up | 2.092249 | SOCS3 |
| 206363_at | 9.68E-06 | down | 2.261549 | MAF |
| 208991_at | 1.49E-13 | down | 3.223685 | STAT3 |
| 209604_s_at | 2.74E-19 | down | 6.546836 | GATA3 |
| 209969_s_at | 2.12E-08 | down | 4.748202 | STAT1 |
| 210426_x_at | 1.14E-12 | down | 6.358444 | RORA |
| 210479_s_at | 5.21E-15 | down | 7.850211 | RORA |
| 212501_at | 1.73E-07 | up | 2.169912 | CEBPB |
| 212549_at | 7.00E-12 | up | 2.369991 | STAT5B |
| 212550_at | 7.19E-10 | up | 2.522143 | STAT5B |
| 213006_at | 6.03E-10 | up | 4.206962 | CEBPD |
| 218221_at | 1.49E-11 | up | 2.349603 | ARNT |
| 218559_s_at | 9.49E-07 | up | 3.354582 | MAFB |
| 218880_at | 5.34E-11 | up | 3.750719 | FOSL2 |
| 208808_s_at | 1.07E-11 | up | 9.120556 | HMGB |
| 200989_at | 1.17E-06 | up | 2.997033 | HIF1A |

Table 7. Cytokine

| Probe ID | Pvalue | Arrow | Fold | Gene |
|---|---|---|---|---|
| 201108_s_at | 2.84E-06 | up | 2.858913 | THBS1 |
| 201109_s_at | 3.87E-05 | up | 3.674066 | THBS1 |
| 201110_s_at | 2.02E-09 | up | 8.271206 | THBS1 |
| 203085_s_at | 1.57E-08 | up | 2.327314 | TGFB1 |
| 203828_s_at | 7.88E-05 | down | 2.130993 | IL32 |
| 205016_at | 8.33E-10 | up | 4.855893 | TGFA |
| 205992_s_at | 4.40E-06 | up | 3.5756 | IL15 |
| 208114_s_at | 7.75E-20 | down | 5.849659 | ISG20L2 |
| 208200_at | 3.06E-11 | down | 4.80094 | IL1A |
| 212195_at | 3.90E-06 | up | 2.667476 | IL6ST |
| 206488_s_at | 1.04E-04 | up | 2.926877 | CD36 |
| 209555_s_at | 2.87E-05 | up | 3.18128 | CD36 |
| 212657_s_at | 2.96E-07 | up | 2.31195 | IL1RN |

Table 8. Cytokine receptor

| Probe ID | Pvalue | Arrow | Fold | Gene |
|---|---|---|---|---|
| 201642_at | 1.42E-09 | up | 2.315 | IFNGR2 |
| 202727_s_at | 1.44E-08 | up | 3.323753 | IFNGR1 |
| 202948_at | 5.77E-10 | up | 6.463163 | IL1R1 |
| 203233_at | 2.36E-10 | up | 3.270304 | IL4R |
| 204191_at | 2.98E-07 | up | 2.05704 | IFNAR1 |
| 204731_at | 7.48E-21 | down | 11.93166 | TGFBR3 |
| 204786_s_at | 5.23E-19 | down | 6.864011 | IFNAR2 |
| 205227_at | 2.89E-05 | up | 2.68359 | IL1RAP |
| 205291_at | 2.89E-08 | down | 2.442178 | IL2RB |
| 205403_at | 1.87E-08 | up | 6.689801 | IL1R2 |
| 205707_at | 1.73E-09 | down | 2.408515 | IL17RA |
| 205798_at | 2.48E-24 | down | 31.78504 | IL7R |
| 205926_at | 1.06E-09 | down | 2.187688 | IL27RA |
| 205945_at | 1.49E-22 | down | 16.68902 | IL6R |
| 206618_at | 4.70E-09 | up | 12.92154 | IL18R1 |
| 207072_at | 5.22E-08 | up | 4.927116 | IL18RAP |
| 211372_s_at | 1.76E-08 | up | 10.6815 | IL1R2 |
| 211676_s_at | 6.66E-09 | up | 4.607373 | IFNGR1 |
| 217489_s_at | 2.79E-14 | down | 3.546462 | IL6R |

Table 9. CSF

| Probe ID | Pvalue | Arrow | Fold | Gene |
|---|---|---|---|---|
| 205159_at | 1.17E-06 | up | 2.511396 | CSF2RB |
| 210340_s_at | 4.36E-10 | up | 2.295372 | CSF2RA |

Table 10. TNF

| Probe ID | Pvalue | Arrow | Fold | Gene |
|---|---|---|---|---|
| 202509_s_at | 1.99E-12 | down | 2.4938 | TNFAIP2 |
| 206026_s_at | 7.48E-06 | up | 3.620447 | TNFAIP6 |
| 206222_at | 1.55E-06 | down | 2.076582 | TNFRSF10C |
| 207536_s_at | 5.06E-07 | down | 2.86682 | TNFRSF9 |
| 207643_s_at | 3.72E-12 | up | 2.681254 | TNFRSF1A |
| 207907_at | 9.40E-17 | down | 3.895788 | TNFSF14 |
| 208296_x_at | 2.21E-05 | up | 2.431529 | TNFAIP8 |
| 210260_s_at | 4.11E-05 | up | 2.505209 | TNFAIP8 |
| 214329_x_at | 1.19E-04 | up | 2.324046 | TNFSF10 |

Table 11. Chemokine

| Probe ID | Pvalue | Arrow | Fold | Gene |
|---|---|---|---|---|
| 200660_at | 9.23E-12 | up | 2.053883 | S100A11 |
| 202917_s_at | 2.79E-12 | up | 2.617102 | S100A8 |
| 203535_at | 3.98E-16 | up | 2.584441 | S100A9 |
| 204103_at | 3.51E-09 | down | 2.364728 | CCL4 |
| 204351_at | 7.20E-04 | up | 2.155134 | S100P |
| 205099_s_at | 6.98E-05 | down | 2.455516 | CCR1 |
| 205863_at | 7.49E-14 | up | 4.166398 | S100A12 |
| 205898_at | 6.76E-04 | down | 2.496165 | CX3CR1 |
| 206337_at | 5.73E-08 | down | 5.099034 | CCR7 |
| 206366_x_at | 1.36E-09 | down | 3.847104 | XCL1 |
| 206978_at | 7.29E-05 | up | 2.362277 | CCR2 |
| 207165_at | 6.52E-05 | up | 2.285972 | HMMR |
| 208304_at | 4.88E-05 | down | 3.621386 | CCR3 |
| 214370_at | 5.04E-06 | down | 2.084171 | S100A8 |
| 214567_s_at | 8.06E-08 | down | 2.902829 | XCL1 /// XCL2 |
| 221058_s_at | 1.49E-09 | up | 2.15809 | CKLF |

Table 12. PGD LTX

| Probe ID | Pvalue | Arrow | Fold | Gene |
|---|---|---|---|---|
| 203913_s_at | 1.26E-08 | up | 16.54476 | HPGD |
| 203914_x_at | 3.77E-08 | up | 14.66064 | HPGD |
| 204445_s_at | 2.84E-06 | up | 2.076126 | ALOX5 |
| 204446_s_at | 1.01E-07 | up | 2.0434 | ALOX5 |
| 204748_at | 0.019719 | up | 2.109052 | PTGS2 |
| 205128_x_at | 5.85E-07 | up | 2.664767 | PTGS1 |
| 207206_s_at | 1.59E-04 | up | 2.531166 | ALOX12 |
| 209533_s_at | 3.14E-10 | up | 2.619734 | PLAA |
| 210128_s_at | 9.02E-10 | up | 2.505719 | LTB4R |
| 210145_at | 1.51E-12 | up | 3.476688 | PLA2G4A |
| 211548_s_at | 4.38E-08 | up | 12.3211 | HPGD |
| 211549_s_at | 3.53E-04 | up | 2.6726 | HPGD |
| 211748_x_at | 1.56E-06 | down | 2.033312 | PTGDS |
| 214366_s_at | 5.17E-10 | up | 3.681805 | ALOX5 |
| 215813_s_at | 1.01E-08 | up | 3.363031 | PTGS1 |
| 215894_at | 2.35E-14 | down | 10.40363 | PTGDR |
| 216388_s_at | 9.35E-07 | up | 2.104241 | LTB4R |
| 208771_s_at | 4.07E-08 | up | 2.455806 | LTA4H |

Table 13. Acute Response Protein

| Probe ID | Pvalue | Arrow | Fold | Gene |
|---|---|---|---|---|
| 200602_at | 3.75E-12 | up | 4.384286 | APP |
| 206157_at | 8.31E-08 | up | 3.272863 | PTX3 |
| 208248_x_at | 2.42E-09 | up | 2.54326 | APLP2 |
| 208691_at | 0.001264 | up | 2.485756 | TFRC |
| 208702_x_at | 7.55E-09 | up | 2.826174 | APLP2 |
| 208703_s_at | 1.26E-07 | up | 3.047052 | APLP2 |
| 208704_x_at | 1.61E-08 | up | 2.435629 | APLP2 |
| 211404_s_at | 4.34E-10 | up | 2.927751 | APLP2 |
| 214875_x_at | 1.32E-08 | up | 2.761566 | APLP2 |
| 214953_s_at | 8.93E-05 | up | 2.120433 | APP |
| 219890_at | 1.43E-12 | up | 7.827181 | CLEC5A |
| 220496_at | 2.59E-07 | up | 3.327139 | CLEC1B |
| 205033_s_at | 1.17E-05 | up | 4.788064 | DEFA1/A1B/A3 |
| 207269_at | 2.87E-05 | up | 6.665461 | DEFA4 |

Table 14. Complement

| Probe ID | Pvalue | Arrow | Fold | Gene |
|---|---|---|---|---|
| 200983_x_at | 7.97E-09 | up | 3.370908 | CD59 |
| 200984_s_at | 9.06E-10 | up | 3.891589 | CD59 |
| 200985_s_at | 4.85E-11 | up | 6.593943 | CD59 |
| 201925_s_at | 2.14E-07 | up | 5.613841 | CD55 |
| 201926_s_at | 6.74E-09 | up | 3.830297 | CD55 |
| 202953_at | 7.01E-06 | up | 2.526228 | C1QB |
| 205786_s_at | 5.02E-13 | up | 4.053864 | ITGAM |
| 206244_at | 6.06E-12 | up | 6.759067 | CR1 |
| 208783_s_at | 0.004769 | up | 2.21095 | CD46 |
| 209906_at | 7.48E-09 | up | 4.336492 | C3AR1 |
| 210184_at | 1.17E-06 | up | 2.086657 | ITGAX |
| 212463_at | 2.34E-09 | up | 2.845059 | CD59 |
| 217552_x_at | 5.04E-10 | up | 3.57143 | CR1 |
| 218232_at | 1.52E-08 | up | 3.972673 | C1QA |
| 218983_at | 7.83E-08 | up | 2.636687 | C1RL |
| 220088_at | 9.13E-08 | up | 2.491036 | C5AR1 |
| 202910_s_at | 3.42E-07 | up | 2.255245 | CD97 |

Table 15. NO/ NADPH oxidase

| Probe ID | Pvalue | Arrow | Fold | Gene |
|---|---|---|---|---|
| 201940_at | 9.58E-11 | up | 5.886338 | CPD |
| 201941_at | 1.14E-09 | up | 5.264568 | CPD |
| 201942_s_at | 6.35E-08 | up | 3.362135 | CPD |
| 201943_s_at | 7.91E-12 | up | 6.937615 | CPD |
| 204961_s_at | 7.26E-08 | up | 2.016737 | NCF1/1B/1C |
| 207677_s_at | 5.88E-10 | up | 2.661943 | NCF4 |
| 214084_x_at | 1.31E-08 | up | 2.251172 | NCF1C |

Table 16. MMP

| Probe ID | Pvalue | Arrow | Fold | Gene |
|---|---|---|---|---|
| 203167_at | 1.02E-13 | up | 3.135463 | TIMP2 |
| 203936_s_at | 2.89E-16 | up | 10.59129 | MMP9 |
| 206871_at | 1.04E-06 | up | 5.3948 | ELANE |
| 207329_at | 3.41E-11 | up | 32.06008 | MMP8 |
| 207890_s_at | 1.30E-11 | up | 3.108211 | MMP25 |
| 202833_s_at | 2.83E-09 | up | 2.778271 | SERPINA1 |
| 204614_at | 5.64E-08 | up | 3.074429 | SERPINB2 |
| 212268_at | 8.64E-11 | up | 5.643009 | SERPINB1 |
| 213572_s_at | 7.52E-11 | up | 5.130388 | SERPINB1 |

Table 17. Caspase

| Probe ID | Pvalue | Arrow | Fold | Gene |
|---|---|---|---|---|
| 202763_at | 1.22E-06 | up | 2.523658 | CASP3 |
| 204780_s_at | 1.71E-04 | up | 2.524168 | FAS |
| 207500_at | 4.39E-06 | up | 2.385929 | CASP5 |
| 208485_x_at | 1.73E-08 | down | 2.568559 | CFLAR |
| 209508_x_at | 1.14E-10 | down | 2.749778 | CFLAR |
| 210564_x_at | 1.77E-07 | down | 2.357083 | CFLAR |
| 210907_s_at | 7.22E-06 | up | 2.816621 | PDCD10 |
| 211316_x_at | 3.60E-13 | down | 3.914512 | CFLAR |
| 211317_s_at | 1.10E-07 | down | 2.657199 | CFLAR |
| 211367_s_at | 7.39E-06 | up | 2.2941 | CASP1 |
| 211862_x_at | 3.34E-08 | down | 2.575973 | CFLAR |
| 213596_at | 4.92E-09 | up | 3.013394 | CASP4 |
| 214486_x_at | 3.71E-08 | down | 2.177017 | CFLAR |
| 215719_x_at | 2.24E-06 | up | 3.678086 | FAS |
| 221601_s_at | 1.49E-09 | down | 4.178467 | FAIM3 |
| 221602_s_at | 2.73E-12 | down | 3.899111 | FAIM3 |

Table 18. Fc receptor

| Probe ID | Pvalue | Arrow | Fold | Gene |
|---|---|---|---|---|
| 203561_at | 6.67E-09 | up | 2.023942 | FCGR2A |
| 204232_at | 8.21E-11 | up | 2.713677 | FCER1G |
| 207674_at | 8.78E-08 | up | 6.420722 | FCAR |
| 210992_x_at | 1.24E-07 | up | 2.313229 | FCGR2C |
| 211307_s_at | 8.29E-08 | up | 4.563443 | FCAR |
| 211395_x_at | 1.20E-06 | up | 2.139768 | FCGR2C |
| 211734_s_at | 3.84E-05 | down | 3.85882 | FCER1A |
| 211816_x_at | 3.58E-05 | up | 2.405171 | FCAR |
| 214511_x_at | 2.07E-05 | up | 3.075548 | FCGR1B |
| 216950_s_at | 5.51E-08 | up | 5.08392 | FCGR1A/1C |

Table 19. CD molecule

| Probe ID | Pvalue | Arrow | Fold | Gene |
|---|---|---|---|---|
| 200663_at | 7.97E-10 | up | 2.446524 | CD63 |
| 201005_at | 4.83E-08 | up | 4.230153 | CD9 |
| 202351_at | 5.59E-10 | up | 3.163429 | ITGAV |
| 202638_s_at | 7.74E-08 | up | 2.891012 | ICAM1 |
| 202878_s_at | 8.07E-06 | up | 2.265542 | CD93 |
| 202910_s_at | 3.42E-07 | up | 2.255245 | CD97 |
| 203645_s_at | 2.02E-09 | up | 9.129274 | CD163 |
| 204306_s_at | 2.48E-06 | up | 2.098005 | CD151 |
| 204489_s_at | 2.80E-09 | up | 2.832196 | CD44 |
| 204490_s_at | 3.30E-09 | up | 2.773283 | CD44 |
| 204661_at | 2.08E-04 | down | 2.102266 | CD52 |
| 205173_x_at | 8.14E-08 | up | 3.565981 | CD58 |
| 205789_at | 6.34E-06 | up | 3.14233 | CD1D |
| 205831_at | 4.40E-10 | down | 3.924635 | CD2 |
| 205988_at | 3.64E-19 | down | 5.606748 | CD84 |
| 206488_s_at | 1.04E-04 | up | 2.926877 | CD36 |
| 206761_at | 5.72E-06 | down | 2.026305 | CD96 |
| 208405_s_at | 5.16E-06 | up | 2.167749 | CD164 |
| 208650_s_at | 7.29E-08 | up | 4.591438 | CD24 |
| 208651_x_at | 5.17E-10 | up | 3.761404 | CD24 |
| 208653_s_at | 1.98E-11 | up | 4.511797 | CD164 |
| 208654_s_at | 3.10E-07 | up | 5.153189 | CD164 |
| 209555_s_at | 2.87E-05 | up | 3.18128 | CD36 |
| 209771_x_at | 1.91E-08 | up | 4.956121 | CD24 |
| 209835_x_at | 3.82E-07 | up | 2.377499 | CD44 |
| 210031_at | 4.04E-09 | down | 3.14224 | CD247 |
| 211744_s_at | 7.96E-09 | up | 3.998247 | CD58 |
| 211900_x_at | 3.47E-14 | down | 2.437045 | CD6 |
| 211945_s_at | 2.07E-06 | up | 2.577267 | ITGB1 |
| 212014_x_at | 4.13E-07 | up | 2.48835 | CD44 |
| 212063_at | 5.59E-07 | up | 2.205469 | CD44 |
| 213958_at | 2.29E-08 | down | 2.119745 | CD6 |
| 215049_x_at | 6.80E-09 | up | 8.964883 | CD163 |
| 216233_at | 3.21E-06 | up | 4.34145 | CD163 |
| 216379_x_at | 6.81E-09 | up | 5.765379 | CD24 |
| 216942_s_at | 4.88E-06 | up | 3.031317 | CD58 |
| 217523_at | 4.41E-13 | down | 6.665958 | CD44 |
| 219669_at | 1.40E-13 | up | 34.68958 | CD177 |
| 222061_at | 6.85E-09 | up | 3.64802 | CD58 |
| 222292_at | 7.05E-11 | down | 2.150798 | CD40 |
| 266_s_at | 1.78E-10 | up | 6.956197 | CD24 |

Table 20. Coagulation

| Probe ID | Pvalue | Arrow | Fold | Gene |
|---|---|---|---|---|
| 203305_at | 2.16E-04 | up | 2.180403 | F13A1 |
| 204714_s_at | 1.87E-08 | up | 3.933558 | F5 |
| 205756_s_at | 2.79E-05 | up | 2.08411 | F8 |
| 205871_at | 7.54E-07 | down | 3.123419 | PLGLA/B1/B2 |
| 206655_s_at | 2.25E-08 | up | 5.369561 | GP1BB/SEPT5 |
| 207808_s_at | 6.30E-08 | up | 2.883896 | PROS1 |
| 210845_s_at | 2.55E-07 | up | 2.502571 | PLAUR |
| 211924_s_at | 5.53E-07 | up | 2.325629 | PLAUR |
| 212245_at | 6.18E-07 | up | 2.301938 | MCFD2 |
| 213258_at | 1.07E-06 | up | 2.352817 | TFPI |
| 213506_at | 0.002877 | up | 2.349815 | F2RL1 |
| 214415_at | 1.30E-09 | down | 5.536361 | PLGLB1/B2 |
| 214866_at | 5.37E-10 | up | 2.031086 | PLAUR |
| 216956_s_at | 4.64E-05 | up | 2.39087 | ITGA2B |
| 218718_at | 2.79E-10 | up | 9.385749 | PDGFC |
| 204627_s_at | 1.30E-06 | up | 4.180416 | ITGB3 |
| 203887_s_at | 8.66E-09 | up | 4.530585 | THBD |
| 203888_at | 4.42E-08 | up | 2.810682 | THBD |

Table 21. Glycolysis

| Probe ID | Pvalue | Arrow | Fold | Gene |
|---|---|---|---|---|
| 200650_s_at | 2.02E-09 | up | 2.710678 | LDHA |
| 200737_at | 2.94E-11 | up | 3.17309 | PGK1 |
| 201030_x_at | 9.45E-05 | down | 2.016562 | LDHB |
| 201251_at | 2.51E-10 | up | 2.67251 | PKM2 |
| 202464_s_at | 6.45E-09 | up | 7.300454 | PFKFB3 |
| 202934_at | 9.80E-14 | up | 4.768903 | HK2 |
| 202990_at | 2.15E-12 | up | 4.196534 | PYGL |
| 203502_at | 1.24E-04 | up | 3.670577 | BPGM |
| 205936_s_at | 5.17E-12 | up | 4.987516 | HK3 |
| 206348_s_at | 9.53E-11 | up | 2.597892 | PDK3 |
| 208308_s_at | 3.92E-09 | up | 2.215685 | GPI |
| 209992_at | 3.99E-09 | up | 11.77066 | PFKFB2 |
| 213453_x_at | 2.13E-12 | up | 2.175151 | GAPDH |
| 217294_s_at | 3.28E-06 | up | 2.62132 | ENO1 |
| 217356_s_at | 4.20E-08 | up | 2.028929 | PGK1 |
| 218273_s_at | 1.01E-07 | down | 2.250674 | PDP1 |

Table 22. H-ATPase

| Probe ID | Pvalue | Arrow | Fold | Gene |
|---|---|---|---|---|
| 200078_s_at | 6.15E-13 | up | 2.530917 | ATP6V0B |
| 201171_at | 4.49E-10 | up | 2.484021 | ATP6V0E1 |
| 201443_s_at | 5.84E-06 | up | 2.32877 | ATP6AP2 |
| 201971_s_at | 4.45E-13 | down | 5.207561 | ATP6V1A |
| 202872_at | 1.95E-10 | up | 6.183733 | ATP6V1C1 |
| 202874_s_at | 6.99E-10 | up | 5.718367 | ATP6V1C1 |
| 204158_s_at | 5.14E-08 | up | 2.068726 | TCIRG1 |
| 208898_at | 2.66E-09 | up | 2.413653 | ATP6V1D |
| 213587_s_at | 1.13E-08 | down | 2.067119 | ATP6V0E2 |
| 206208_at | 1.00E-11 | up | 3.51149 | CA4 |
| 206209_s_at | 4.18E-15 | up | 7.982899 | CA4 |
| 209301_at | 2.78E-06 | up | 3.422036 | CA2 |
| 212536_at | 4.38E-09 | up | 4.21056 | ATP11B |
| 213582_at | 1.89E-08 | up | 2.241957 | ATP11A |

Table 23. Vasodilator

| Probe ID | Pvalue | Arrow | Fold | Gene |
|---|---|---|---|---|
| 201494_at | 1.01E-08 | up | 2.190291 | PRCP |
| 202912_at | 1.20E-08 | up | 4.330455 | ADM |
| 212741_at | 0.004196 | up | 2.027247 | MAOA |

Table 24. NK cell

| Probe ID | Pvalue | Arrow | Fold | Gene |
|---|---|---|---|---|
| 202379_s_at | 2.01E-30 | down | 26.43984 | NKTR |
| 205821_at | 1.39E-12 | down | 3.985353 | KLRK1 |
| 207509_s_at | 7.71E-10 | down | 2.948439 | LAIR2 |
| 207795_s_at | 1.01E-09 | down | 3.216148 | KLRD1 |
| 210288_at | 1.62E-11 | down | 5.382525 | KLRG1 |
| 210606_x_at | 1.91E-09 | down | 3.341239 | KLRD1 |
| 214470_at | 1.11E-04 | down | 2.103258 | KLRB1 |
| 215338_s_at | 1.06E-24 | down | 14.54682 | NKTR |
| 220646_s_at | 1.34E-04 | down | 2.386466 | KLRF1 |
| 205488_at | 1.01E-05 | down | 2.867692 | GZMA |
| 206666_at | 1.84E-07 | down | 3.446082 | GZMK |
| 207460_at | 3.78E-09 | down | 2.502287 | GZMM |
| 210164_at | 8.91E-09 | down | 3.75597 | GZMB |
| 210321_at | 8.94E-10 | down | 5.800327 | GZMH |
| 214617_at | 2.22E-06 | down | 2.646147 | PRF1 |

Table 25. T cell

| Probe ID | Pvalue | Arrow | Fold | Gene |
|---|---|---|---|---|
| 205039_s_at | 2.79E-08 | down | 2.22558 | IKZF1 |
| 205255_x_at | 3.09E-08 | down | 2.955244 | TCF7 |
| 205456_at | 5.31E-08 | down | 2.877146 | CD3E |
| 205488_at | 1.01E-05 | down | 2.867692 | GZMA |
| 205495_s_at | 5.33E-10 | down | 4.378694 | GNLY |
| 205758_at | 1.20E-07 | down | 3.258815 | CD8A |
| 206666_at | 1.84E-07 | down | 3.446082 | GZMK |
| 206804_at | 1.10E-15 | down | 5.118528 | CD3G |
| 207460_at | 3.78E-09 | down | 2.502287 | GZMM |
| 208003_s_at | 5.52E-18 | down | 12.03963 | NFAT5 |
| 209670_at | 5.21E-06 | down | 2.475029 | TRAC |
| 209671_x_at | 3.58E-08 | down | 2.774547 | TRAC |
| 209813_x_at | 1.49E-09 | down | 4.424708 | TARP |
| 210164_at | 8.91E-09 | down | 3.75597 | GZMB |
| 210321_at | 8.94E-10 | down | 5.800327 | GZMH |
| 210370_s_at | 1.34E-07 | down | 2.482685 | LY9 |
| 210555_s_at | 1.02E-07 | down | 2.476832 | NFATC3 |
| 210556_at | 4.68E-08 | down | 2.850907 | NFATC3 |
| 210915_x_at | 6.23E-06 | down | 2.847533 | TRBC1 |
| 210972_x_at | 1.78E-07 | down | 2.875805 | TRAC/J17/V20 |
| 211144_x_at | 5.76E-08 | down | 3.696107 | TARP/TRGC2 |
| 211796_s_at | 6.35E-06 | down | 2.926207 | TRBC1/C2 |
| 211902_x_at | 8.99E-07 | down | 2.286838 | TRD@ |
| 212759_s_at | 3.98E-16 | down | 3.926594 | TCF7L2 |
| 212762_s_at | 2.70E-09 | down | 2.376013 | TCF7L2 |
| 212808_at | 1.21E-21 | down | 5.549449 | NFATC2IP |
| 213193_x_at | 2.53E-06 | down | 2.918569 | TRBC1 |
| 213539_at | 1.00E-08 | down | 3.193378 | CD3D |
| 213830_at | 5.93E-08 | down | 3.51036 | TRD@ |
| 214617_at | 2.22E-06 | down | 2.646147 | PRF1 |
| 215092_s_at | 1.36E-09 | down | 2.475134 | NFAT5 |
| 215806_x_at | 1.02E-08 | down | 4.078028 | TARP/TRGC2 |
| 216191_s_at | 4.71E-07 | down | 4.762182 | TRDV3 |
| 216920_s_at | 2.28E-10 | down | 5.341667 | TARP/TRGC2 |
| 217143_s_at | 1.26E-08 | down | 6.055404 | TRD@ |
| 217526_at | 1.43E-12 | down | 3.846013 | NFATC2IP |
| 217527_s_at | 2.12E-13 | down | 5.801224 | NFATC2IP |
| 220684_at | 7.39E-09 | down | 2.077709 | TBX21 |
| 220704_at | 2.15E-10 | down | 5.686157 | IKZF1 |
| 37145_at | 9.67E-10 | down | 4.340701 | GNLY |
| 214032_at | 6.60E-08 | down | 2.523588 | ZAP70 |
| 204890_s_at | 1.65E-07 | down | 2.638662 | LCK |
| 204891_s_at | 4.58E-08 | down | 3.313788 | LCK |
| 205831_at | 4.40E-10 | down | 3.924635 | CD2 |
| 201565_s_at | 8.13E-13 | down | 4.167651 | ID2 |
| 213931_at | 7.33E-08 | down | 3.546731 | ID2/2B |

Table 26. B cell

| Probe ID | Pvalue | Arrow | Fold | Gene |
|---|---|---|---|---|
| 221969_at | 9.96E-13 | down | 4.199424 | PAX5 |
| 203140_at | 3.09E-10 | up | 3.687249 | BCL6 |
| 210105_s_at | 9.37E-10 | down | 3.319013 | FYN |
| 210754_s_at | 2.98E-10 | down | 3.545109 | LYN |
| 205039_s_at | 2.79E-08 | down | 2.22558 | IKZF1 |
| 211430_s_at | 0.015735 | up | 2.830679 | IGHG1/G2 |
| 211643_x_at | 0.024398 | up | 2.10616 | IGK |
| 212592_at | 0.017267 | up | 2.569312 | IGJ |
| 212827_at | 0.008027 | down | 2.240154 | IGHM |
| 214677_x_at | 0.031357 | up | 2.035694 | IGLV1-44 |
| 214768_x_at | 0.006798 | up | 2.374888 | IGKV1-5 |
| 217022_s_at | 5.28E-05 | up | 5.197489 | IGHA1 /A2 |
| 210970_s_at | 2.98E-06 | up | 2.298244 | IBTK |
| 217620_s_at | 4.31E-12 | down | 2.785986 | PIK3CB |
| 221756_at | 5.52E-10 | down | 2.616271 | PIK3IP1 |
| 204053_x_at | 1.57E-06 | up | 2.560581 | PTEN |
| 204054_at | 2.56E-10 | up | 5.50691 | PTEN |
| 211711_s_at | 2.54E-08 | up | 4.673204 | PTEN |
| 206370_at | 2.95E-09 | down | 2.443374 | PIK3CG |
| 212240_s_at | 4.29E-13 | down | 4.50699 | PIK3R1 |
| 212249_at | 1.96E-06 | down | 2.560644 | PIK3R1 |